%% file: AHCAL_TP_paper.tex
\documentclass[11pt,a4paper]{article}
\usepackage{jinstpub}
\usepackage{graphicx}
\usepackage{epsfig}
\usepackage{url}
\usepackage{fontenc,times}
\usepackage{wrapfig,rotating}
\usepackage{caption}           % extended features for float captions
\usepackage{subcaption} % support for sub-floats

\usepackage[usenames]{color}
\makeatletter\let\default@color\current@color\makeatother
\usepackage{tikz}
\usepackage{lineno}
\title{\begin{center}
Design, Construction and Commissioning of a Technological Prototype of a Highly Granular SiPM-on-tile Scintillator-Steel Hadronic Calorimeter 
\end{center}
\vspace{1cm}
}

\collaboration{\begin{center}The CALICE Collaboration \end{center}}

\subheader{\begin{flushright}CALICE-PUB-2022-003\end{flushright}}

\abstract{
The CALICE collaboration is developing highly granular electromagnetic and hadronic calorimeters for detectors at future energy frontier electron-positron colliders. After successful tests of a physics prototype, a technological prototype of the Analog Hadron Calorimeter has been built, based on a design and construction techniques scalable to a collider detector. The prototype
consists of a steel absorber structure and active layers of small scintillator tiles that are individually read out by directly coupled SiPMs. 
Each layer has an active area of $72 \times 72\,{\rm cm}^{2}$ and a tile size of $3 \times 3\,{\rm cm}^{2}$. With $38$ active layers, the prototype has nearly $22,000$ readout channels, and its total thickness amounts to $4.4$ nuclear interaction lengths. The dedicated readout electronics provide time stamping of each hit with an expected resolution of about~$1\,{\rm ns}$.
The prototype was constructed in 2017 and commissioned in beam tests at DESY. It recorded muons, hadron showers and electron showers at different energies in test beams at CERN in 2018.
In this paper, the design of the prototype, its construction and commissioning are described. The methods used to calibrate the detector are detailed, and the performance achieved in terms of uniformity and stability is presented.
}

\begin{document}
\maketitle
%\begin{frontmatter}

%\tableofcontents
%\linenumbers
%\pagewiselinenumbers

\input{introduction.tex}
\input{prototype.tex}
\input{construction.tex}
\input{commissioning.tex}
\input{calibration_split.tex}

\input{testbeam.tex}

\input{conclusion.tex}

\section*{Acknowledgements}

We would like to thank the technicians and the engineers who contributed to the design and construction of the prototypes. We express our gratitude to the DESY and CERN laboratories for hosting our test beam experiments, and to their staff for the efficient accelerator operation and excellent support. Some of the measurements leading to these results have been performed at the Test Beam Facility at DESY Hamburg (Germany), a member of the Helmholtz Association (HGF).
This work was supported by the FWO, Belgium; by the Natural Sciences and Engineering Research Council of Canada; by the Ministry of Education, Youth and Sports of the Czech Republic; by the AIDA-2020 Advanced European Infrastructures for Detectors at Accelerators project, which has received funding from the European Union’s Horizon 2020 Research and Innovation programme under Grant Agreement no. 654168; by the P2IO LabEx in the framework ’Investissements d’Avenir’ managed by the French National Research Agency (ANR) under Grant Agreements ANR-10-LABX-0038 and ANR-11-IDEX-0003-01; by the Master Project CALICE of CNRS/IN2P3 France; by the Alexander von Humboldt Stiftung (AvH), Germany; by the Bundesministerium für Bildung und Forschung (BMBF), Germany, within the Verbundprojekt 05H2018 - R\&D DETEKTOREN (Szintillatoren): R\&D für künftige Szintillatorbasierte Detektoren (AHCAL); by the Deutsche Forschungsgemeinschaft (DFG), Germany; by the Helmholtz-Gemeinschaft (HGF), Germany; by the I-CORE Program of the Planning and Budgeting Committee, Israel; by the Nella and Leon Benoziyo Center for High Energy Physics, Israel; by the Israeli Science Foundation, Israel; by the JSPS KAKENHI Grant-in-Aid for Scientific Research (B) No. 17340071 and specially promoted research No. 223000002, Japan; by the National Research Foundation of Korea; by the Korea-EU cooperation programme of National Research Foundation of Korea, Grant Agreement 2014K1A3A7A03075053; by the Russian Ministry of Education and Science contracts 3.2989.2017 and 14.W03.31.0026; by the Spanish Ministry of Economy and Competitiveness FPA2014-53938-C3-R2 and Grant MDM-2015-0509; by the Science and Technology Facilities Council, UK; by the US Department of Energy; by the National Science Foundation of the United States of America, and by the Nuclear Physics, Particle Physics, Astrophysics and Cosmology Initiative, a Laboratory Directed Research, USA.

\clearpage

\bibliography{AHCAL_TP_paper}  
\bibliographystyle{JHEP}%\bibliographystyle{plain}

%\appendix
%\input{sections/appendix.tex}
% \section{}
% \label{}

\end{document}

%% file: introduction.tex
\section{Introduction}

The detailed investigation of the properties of the Higgs boson has been identified as the highest priority for particle physics beyond the exploitation of the potential of the High-Luminosity LHC. The best precision for many observables can be obtained in electron-positron collisions, for which a number of different collider options have been proposed: on the one hand linear colliders, such as the ILC~\cite{ref:ILC}, which is under political consideration in Japan, or CLIC~\cite{ref:CLICCDR1, ref:CLICCDR2, ref:CLICCDR3, ref:CLICUpdatedBaseline}, and on the other hand circular colliders such as FCC-ee~\cite{ref:FCCee} or CEPC~\cite{ref:CEPCaccelerator, ref:CEPCdetector}. 

For all these projects, the physics goals place stringent requirements on the detector performance. In particular, the jet energy resolution of $3\%$ to $4\%$ for jet energies up to a few hundred ${\rm GeV}$, which is needed to distinguish hadronically decaying W and Z bosons, cannot be reached by existing particle physics detectors. It has been demonstrated that the Particle Flow Algorithm (PFA)~\cite{ref:Pandora,ref:PFABrient,ref:PFAMorgunov} approach to the reconstruction of jets can achieve this goal for typical jets at ILC energies. In the PFA approach, the energy of each individual particle within a jet is reconstructed in the subdetector with the best energy resolution. One important ingredient for a detector designed for PFA are highly granular calorimeters, which allow an efficient assignment of calorimeter energy depositions caused by charged particles to the respective tracks. Misassignments, called {\em confusion}, lead to a deterioration of the jet energy resolution especially at large jet energies.

The CALICE collaboration
%~\cite{ref:CALICE} 
studies several designs for highly granular calorimeters, for the electromagnetic (ECAL) as well as the hadronic (HCAL) calorimeter. The R\&D on most of these designs proceeds in two phases: in the first phase, {\em physics prototypes} have been built to study the performance reachable with highly granular calorimeters. In the second phase, {\em technological prototypes} focus on engineering and integration solutions needed to include such a calorimeter in a collider detector. 

The technological prototype for the analog hadron calorimeter (AHCAL) described in this document is a high granularity sampling calorimeter based on steel absorber and small scintillator tiles, read out individually by silicon photomultipliers (SiPMs) directly coupled to the tiles. This concept has first been studied with the AHCAL physics prototype~\cite{ref:AHCALPhysProt}.
The calorimeter endcap upgrade of the CMS experiment for the High-Luminosity LHC, the  High-Granularity Calorimeter (HGCAL)~\cite{ref:HGCAL}, incorporates a design following the AHCAL concept in its backward part, where radiation levels allow the use of SiPMs.

%% file: prototype.tex
\section{Design of the AHCAL Technological Prototype}
\label{design}
The design of the AHCAL technological prototype described here is inspired by the calorimeter layout of the International Large Detector concept (ILD)~\cite{ref:ILDLOI,ref:ILCTDRDet,ref:ILDIDR}. Similar calorimeter concepts are discussed for SiD~\cite{ref:SiDLOI,ref:ILCTDRDet}, the Silicon Detector for ILC, and for CLICdet~\cite{ref:CLICdet1,ref:CLICdet2}, the detector model for CLIC, as well as for detector concepts at future circular electron-positron colliders. The AHCAL is constructed as a sampling calorimeter consisting of steel absorber plates and scintillator tiles as the active medium. The scintillator tiles are read out by SiPMs directly coupled to the tiles. This concept is different from most existing tile calorimeters which are read out via long fibers. It allows the integration of both the photosensors and the front-end electronics into the detector volume, making the high granularity required for particle flow methods realizable with scintillators in a very compact and cost-effective way. 

The feasibility of a highly granular scintillator calorimeter concept has been demonstrated with the AHCAL physics prototype~\cite{ref:AHCALPhysProt}. With this cubic-metre sized prototype, the energy resolution for electrons~\cite{ref:AHCALEM} as well as hadrons~\cite{ref:AHCALHad} has been measured, and the detailed structure of hadronic showers~\cite{ref:AHCALTracks, ref:AHCALMCmodels, ref:AHCALPionProton, ref:AHCALDecomposition} has been studied. The separation power for two close-by hadron showers, an important ingredient for particle flow algorithms, has been demonstrated~\cite{ref:AHCALPFA}. The prototype was also successfully tested with tungsten absorbers~\cite{ref:AHCALTungstenLowE, ref:AHCALTungstenHighE}. A consistent hadronic energy resolution has been obtained for a setup with a highly granular scintillator-tungsten electromagnetic calorimeter in front of the AHCAL~\cite{ref:CombinedScint}. 
An overview of the results can be found in~\cite{ref:PFACalorimetry}.

The focus for the design of the AHCAL technological prototype lies on the viability of the concept for a full collider detector: the scalability of the detector layout, taking into account realistic space constraints in a large particle physics detector, the scalability of the production methods, the reliable operation of a large detector and the feasibility of the calibration procedures for a large amount of readout channels. In contrast to the physics prototype, the AHCAL technological prototype also provides information on the time-of-arrival of each hit with an expected resolution of the order of $1\,{\rm ns}$. First studies of the time evolution of hadronic showers have been performed earlier with a small number of scintillator tiles installed behind a large hadron calorimeter prototype~\cite{ref:T3B}. They confirmed the expectation that a time resolution of the order of $1\,{\rm ns}$ gives access to processes developing on different time scales within hadronic showers. The availability of the time information for all channels in a hadron calorimeter prototype opens the way for detailed investigations of hadronic shower shapes and tests of simulation models not only in space and energy, but also time, and to explore the use of detailed time information in the event reconstruction.

\subsection{Design Considerations}
\label{protoPhysCons}

%\subsubsection{Granularity}
The segmentation of the detector is driven mainly by two considerations: the ability to distinguish the energy depositions of close-by neutral and charged hadrons to minimise confusion in the particle flow reconstruction, and the goal to resolve electromagnetic sub-showers in hadron showers for the application of software compensation algorithms. Both require a 
longitudinal segmentation of the order of one radiation length X$_0$ and a transverse segmentation of a size similar to the Moli\`ere radius. Detailed simulation studies of di-jet events in ILD have shown that for a scintillator-steel hadron calorimeter a transverse tile dimension of $3 \times 3\,{\rm cm}^2$ allows a significant improvement of the jet energy resolution by applying software compensation techniques, and that this tile size is an optimal choice considering performance and overall detector channel count~\cite{ref:GranularityJER}. These studies have also shown that a larger tile size of $6 \times 6\,{\rm cm}^2$ in the rear part of the calorimeter could reduce the channel count with only a modest impact on the jet energy resolution.
	      
For the absorber material, non-magnetic stainless steel offers favourable mechanical properties like rigidity and ease of machining, such that a self-supporting structure without auxiliary supports (dead regions) can be realised. Moreover, in contrast to heavier materials, iron has a moderate ratio of hadronic interaction length to electromagnetic radiation length. This  
allows a fine longitudinal sampling in terms of X$_0$ with a reasonable number of layers in a given total hadronic absorption length. These considerations lead to a choice of $1.7\,{\rm cm}$ thick non-magnetic stainless steel absorber layers, corresponding to roughly $1$ radiation length or $0.1$ interaction length.

\begin{figure}[htb]
  \begin{subfigure}[T]{0.5\textwidth}
    \includegraphics[width=\linewidth]{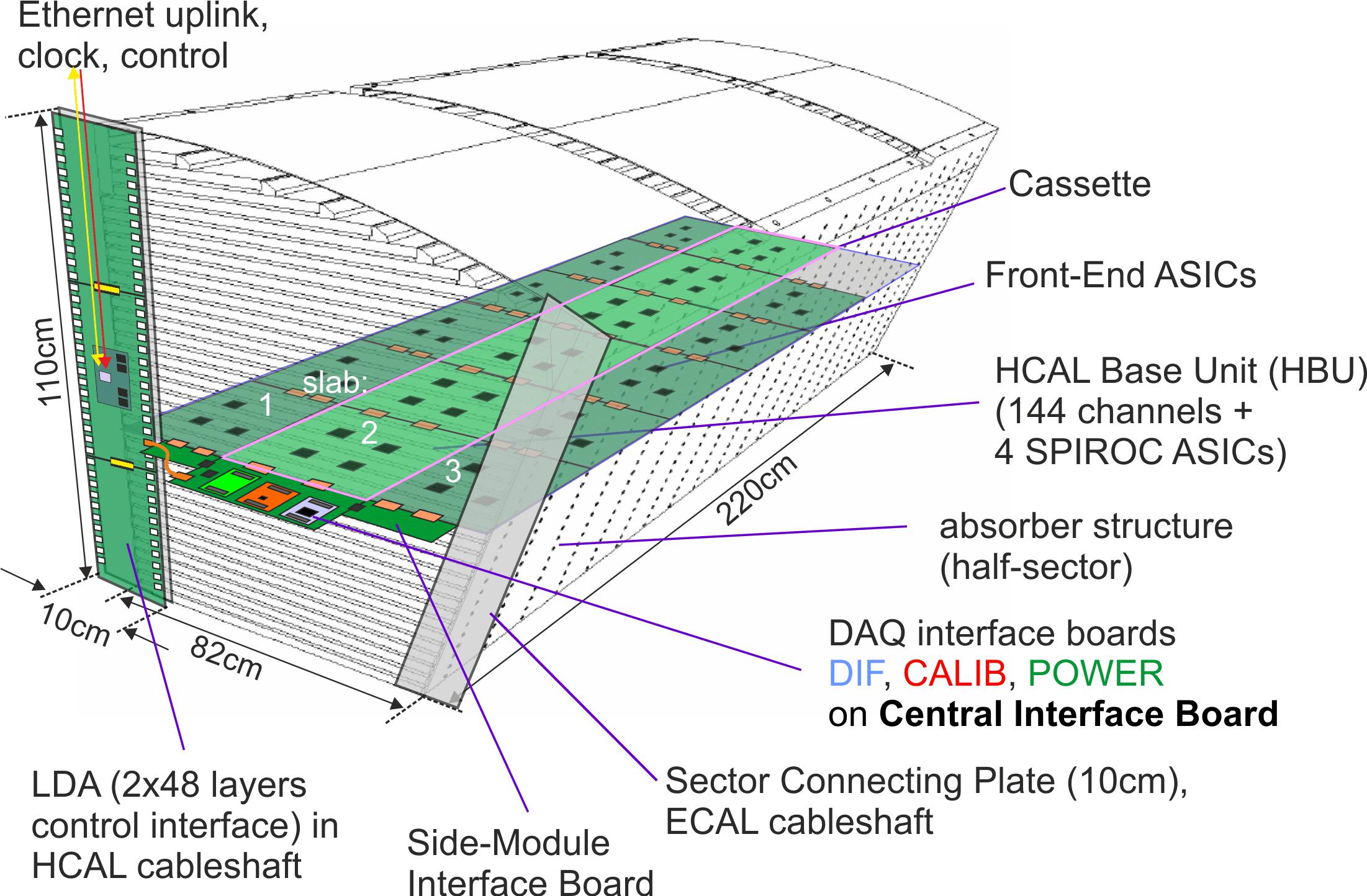}
    \caption{}\label{fig:ILDBarrelSector}
  \end{subfigure}%
  \hfill
  \begin{subfigure}[T]{0.47\textwidth}
    \includegraphics[width=\linewidth]{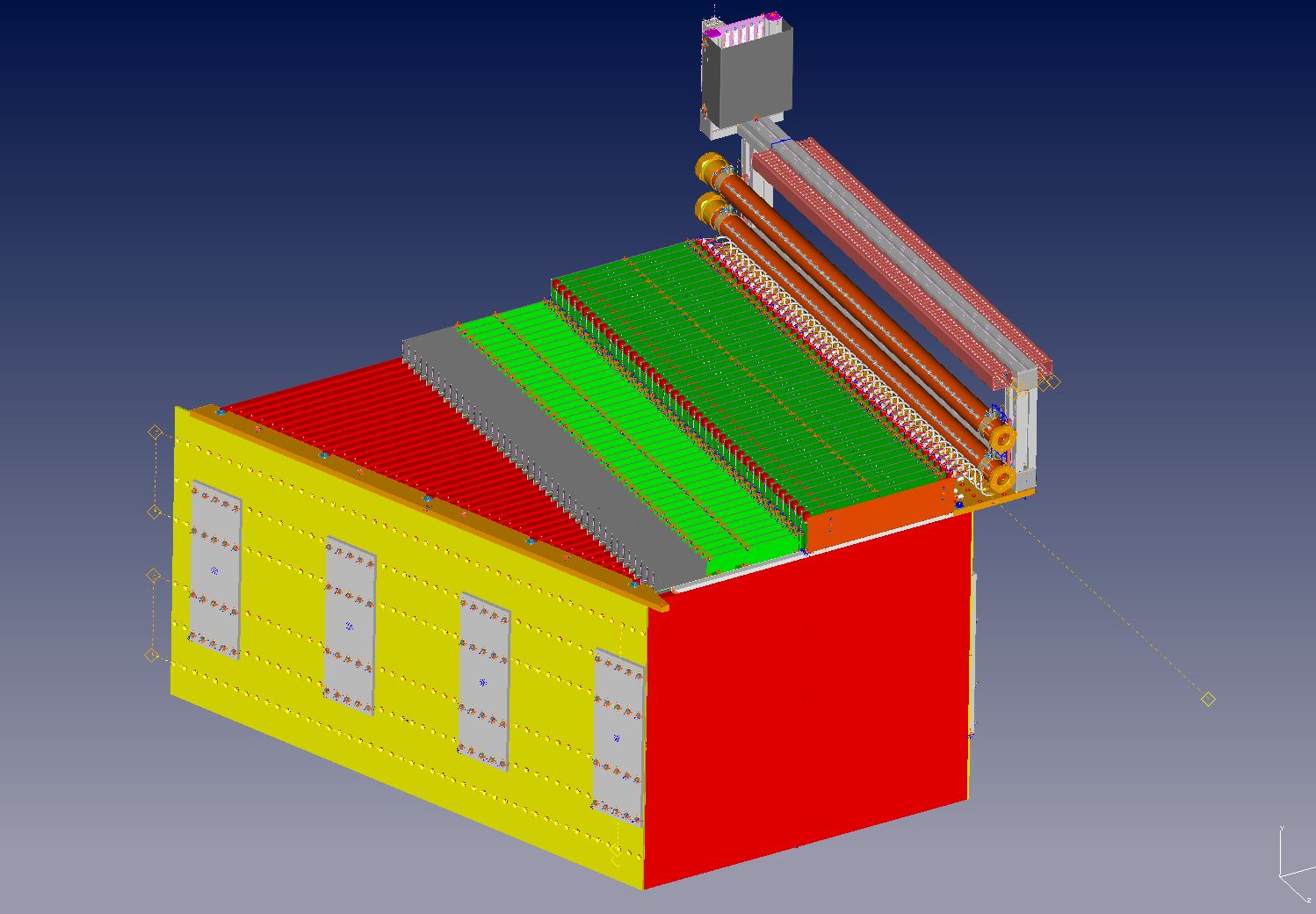}
    \caption{}\label{fig:Absorber}
  \end{subfigure}%
  \caption{Schematic view of (\subref{fig:ILDBarrelSector}) an ILD barrel sector and of (\subref{fig:Absorber}) the testbeam setup of the AHCAL technological prototype. The absorber structure is shown in red (absorber plates) and yellow (connecting plate ensuring the distance between the absorber plates). The green parts correspond to the Central and Side-Module Interface Boards of the active layers, and the orange parts to cooling plates and pipes. A support structure for DAQ boards and cables is indicated in grey. }
    \label{fig:Mechanics}
\end{figure}

In order to ease production and assembly, the layout of the active layers is modular. The basic unit is the HCAL Base Unit (HBU) with a size of $36 \times 36\,{\rm cm}^2$, containing $144$ tiles. Up to six of them can be connected in a row to form a {\it slab}. An active layer is formed by up to three slabs, served by a common set of interface boards providing the connection to power supplies and the Data Acquisition (DAQ) system. This design allows an efficient covering of the different widths of active layers in the barrel sectors of a collider detector (see figure~\ref{fig:ILDBarrelSector}) with only a small number of additional, narrower HBU sizes~\cite{ref:EUDETLayer}. The slabs are contained in individual cassettes for mechanical protection.  

For the AHCAL technological prototype, the active layers consist of four HBUs. These are mounted in {\it modules} formed by two slabs of two HBUs each. They are installed in an absorber structure which has a wedge shape similar to the shape of a sector in the ILD barrel (figure~\ref{fig:Absorber}).
The active layer size of $72 \times 72\,{\rm cm}^2$ is smaller than for the  AHCAL physics prototype with a lateral size of $90 \times 90\,{\rm cm}^2$, leading to a slightly larger lateral leakage for hadron showers.

\subsection{Active Layers}
\label{sub:Active}

\begin{figure}[ht]
  \begin{subfigure}[T]{0.38\textwidth}
     \includegraphics[width=\linewidth]{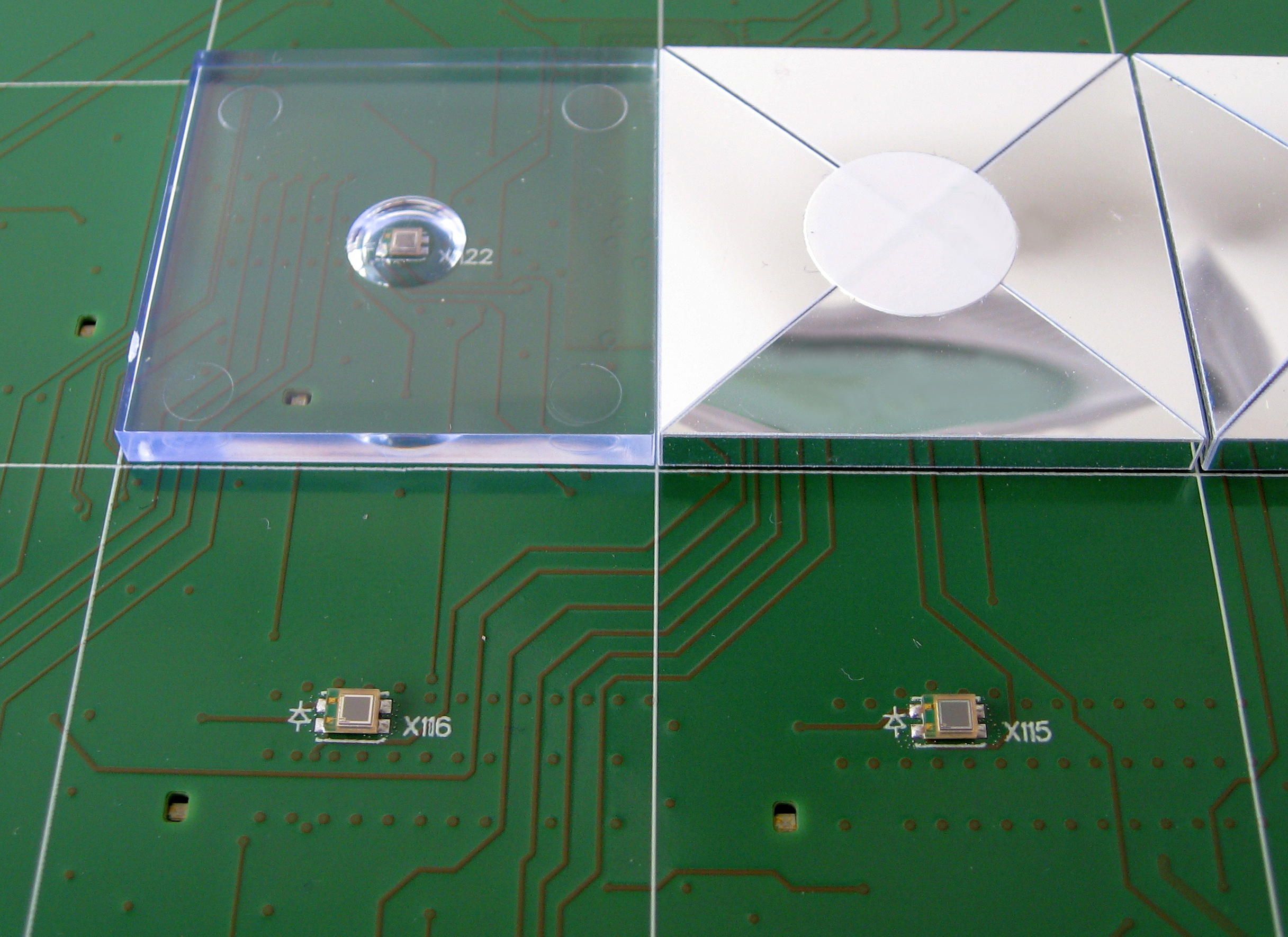}
     \caption{}\label{fig:Tile}
  \end{subfigure}%
  \hfill
  \begin{subfigure}[T]{0.58\textwidth}
     \includegraphics[width=\linewidth]{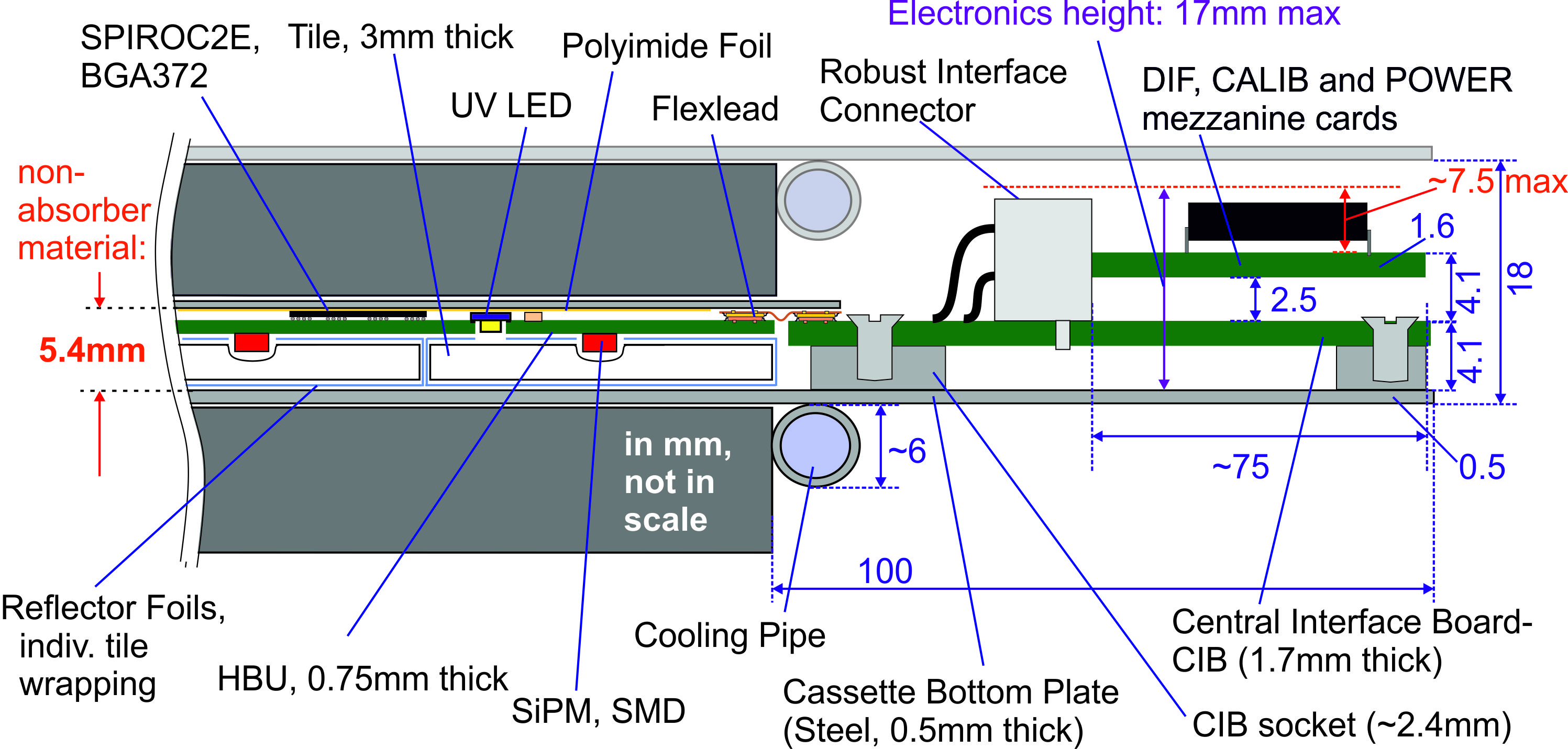}
     \caption{}\label{fig:XSecLayer}
  \end{subfigure}%
  \caption{(\subref{fig:Tile}) Photo of SiPMs on an HBU, two of them with naked and wrapped scintillator tiles. (\subref{fig:XSecLayer}) Schematic view of the cross section of an active layer. }
    \label{fig:ActiveLayer}
\end{figure}

The active material of the AHCAL technological prototype consists of injection-moulded poly\-sty\-rene scintillator tiles with a size of $3 \times 3\,{\rm cm}^2$ and a thickness of $3\,{\rm mm}$. The injection moulding technique is scalable and cost-effective, making it an attractive choice even though the light output is of the order of $30\%$ less than for cast scintillator. No further machining or polishing of the tiles is necessary. A small dimple in the center of the tile (see figure~\ref{fig:Tile})
%, which has been optimised for uniformity and efficiency of the light collection, 
houses the SiPM. This geometry allows the mounting of the SiPM directly onto the electronics boards with standard PCB assembly techniques. A tile design with the SiPM directly coupled to the scintillator tile was originally proposed in~\cite{ref:DirectCouplingNIU} and subsequently varied and optimised in further studies~\cite{ref:DirectCouplingMPI, ref:Dimple}. The signal non-uniformity within a tile and the effects of misalignment of the SiPM with respect to the dimple were found to be small for realistic misalignments~\cite{ref:Uniformity}.
The tiles are individually wrapped in 3M ESR~\cite{ref:ESR} reflective foil (see section~\ref{sub:Wrapping}), leading to negligible optical cross talk between the tiles. 

The system of scintillator tile, reflective foil wrapping and SiPM determines the response of the detector to minimum ionising particles (MIPs). The scintillator material and thickness as well as the SiPM properties are chosen such that the smallest relevant physics signal, given by the energy deposition of a MIP, can be detected with high efficiency, without being too sensitive to SiPM noise. For the design signal size of around $15$ fired pixels per MIP, a threshold of $0.5$ MIP corresponds to $95\%$ or more MIP detection efficiency and leads to a very strong suppression of SiPM noise. The current generation of SiPMs with trenches between pixels has a very low inter-pixel cross talk probability of the order of a few percent, leading to a practically noise-free detector for thresholds as low as $0.25$ to $0.3$ MIP. The total number of pixels of the SiPM determines the maximum energy deposition that can be measured. The saturation of the SiPM due to the finite number of pixels leads to a strongly non-linear behaviour at large amplitudes. In order to measure the expected maximum hit energy of a few hundred MIPs in hadron showers in the ILC energy range, the SiPMs should have of the order of $3,000$ pixels. In addition to these physics considerations, the SiPM properties have to match the characteristics of the readout ASICs. After a tendering process, Hamamatsu MPPCs of type S13360-1325PE~\cite{ref:HPKSiPM_S13360} have been selected. They have a size of $1.3 \times 1.3\,{\rm mm}^2$, with $2668$ pixels with a pitch of $25\,{\rm \mu m}$. Their breakdown voltage is around $53\,{\rm V}$, and they are operated at $5\,{\rm V}$ overvoltage, resulting typically in a photodetection efficiency of $25\%$ and a gain of $7 \times 10^5$. They offer a low dark count rate (of the order of $100\,{\rm kHz}$), very small cross talk ($1\%$) and a small temperature sensitivity ($54\,{\rm mV/K}$). They have been ordered in batches of $600$ devices with breakdown voltages within $\pm 100\,{\rm mV}$, allowing all SiPMs on one module to be operated with the same bias voltage. 

\begin{figure}[htb]
  \begin{subfigure}[T]{0.49\textwidth}
     \includegraphics[width=\linewidth]{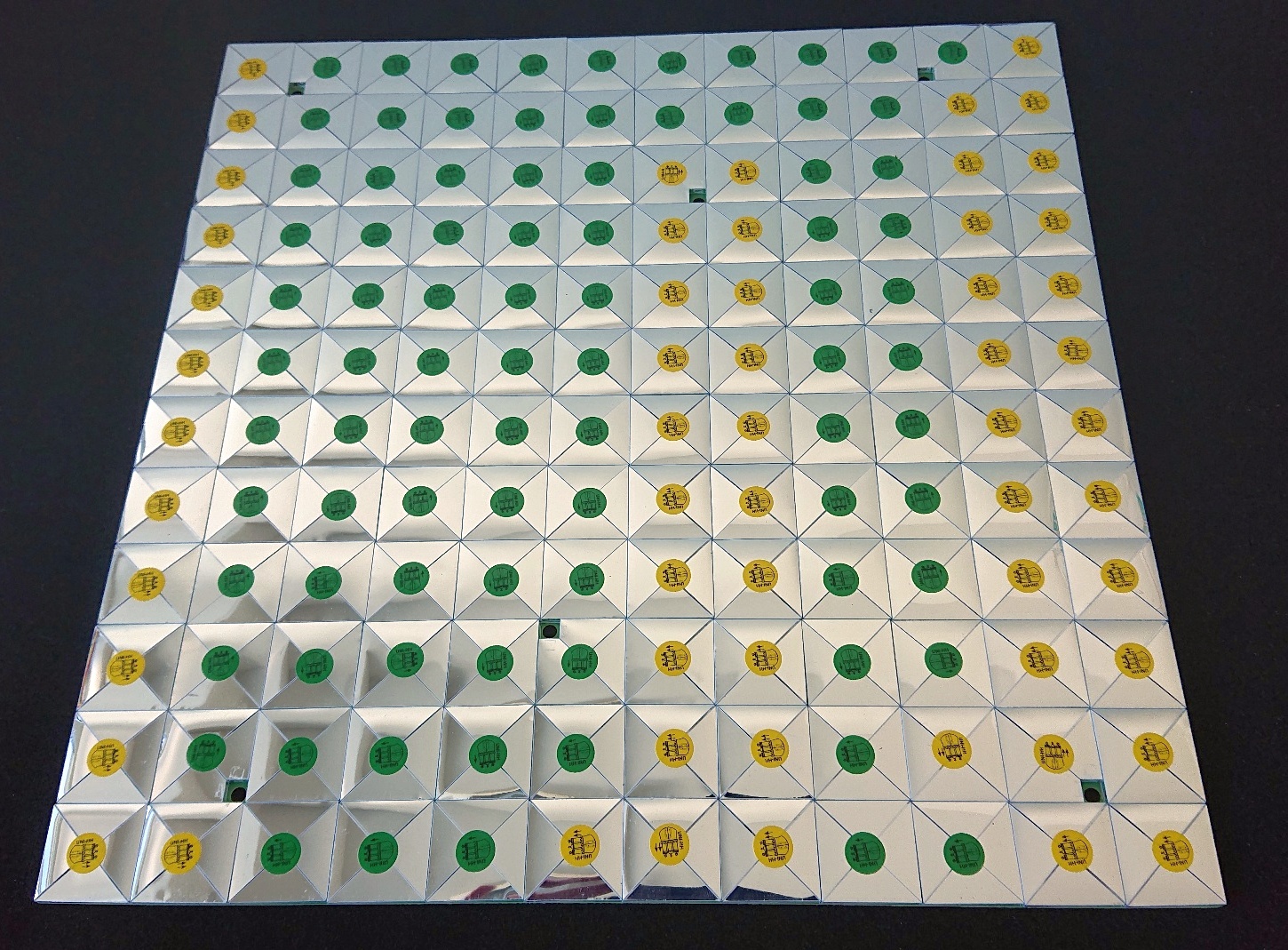}
     \caption{}\label{fig:HBUtop}
  \end{subfigure}%
  \hfill
  \begin{subfigure}[T]{0.49\textwidth}
     \includegraphics[trim=250 20 250 0,clip,width=\linewidth]{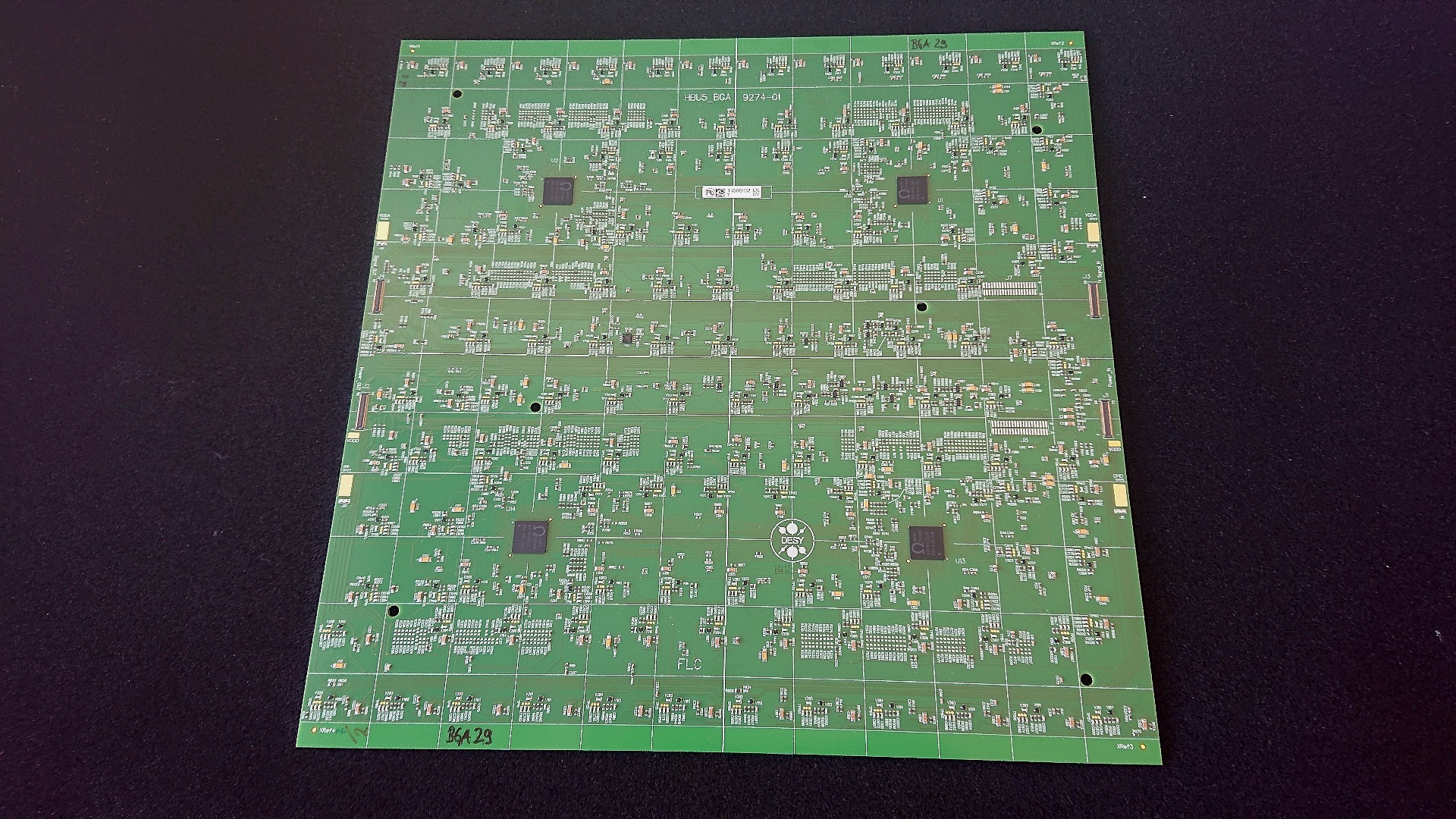}
     \caption{}\label{fig:HBUbottom}
  \end{subfigure}%
  \caption{HBU seen from (\subref{fig:HBUtop}) the top and (\subref{fig:HBUbottom}) the bottom. }
    \label{fig:HBU}
\end{figure}

The $144$ scintillator tiles and SiPMs are mounted on the top side of the HBU, while $4$ SPIROC2E~\cite{ref:SPIROC} readout ASICs as well as other electronics components are located on the bottom side. These include a calibration system with one LED per channel. A thin polyimide foil ($125\,{\mu \rm m}$ thickness) provides the electrical insulation between the PCB and the cassette. A schematic cross section of an active layer is shown in figure~\ref{fig:XSecLayer}. Figure~\ref{fig:HBU} shows photos of the top and bottom sides of a fully assembled HBU.

The electrical connections between the HBUs as well as to the interface board within a slab are realised with very thin flex-leads. One of the slabs has a Central Interface Board (CIB) with active mezzanines (section~\ref{sub:DAQ}), while the other has a passive Side Interface Board (SIB). The slabs are connected with each other with larger flexible connections. 

In total $38$ modules with $3 \times 3\,{\rm cm}^2$ scintillator tiles were constructed. In addition, one module with $6 \times 6\,{\rm cm}^2$ tiles was built. In order to be able to use the same PCBs for the HBUs with $6 \times 6\,{\rm cm}^2$ tiles, the dimples of those larger tiles are not in their centre, but moved to coincide with one of the nominal SiPM positions on the HBU. The lower light output of the larger tiles is compensated by using larger SiPMs. $2 \times 2\,{\rm mm}^2$ SiPMs are built by an array of four Hamamatsu TSV MPPC S13615-1025~\cite{ref:HPKSiPM_S13615}, which have otherwise the same properties (photodetection efficiency, gain, breakdown voltage) as the S13360-1325PE MPPCs.

\subsection{Electronics and Data Acquisition System}
\label{sub:DAQ}
\begin{figure}[htb]
  \begin{subfigure}[T]{0.3\textwidth}
     \includegraphics[width=\linewidth]{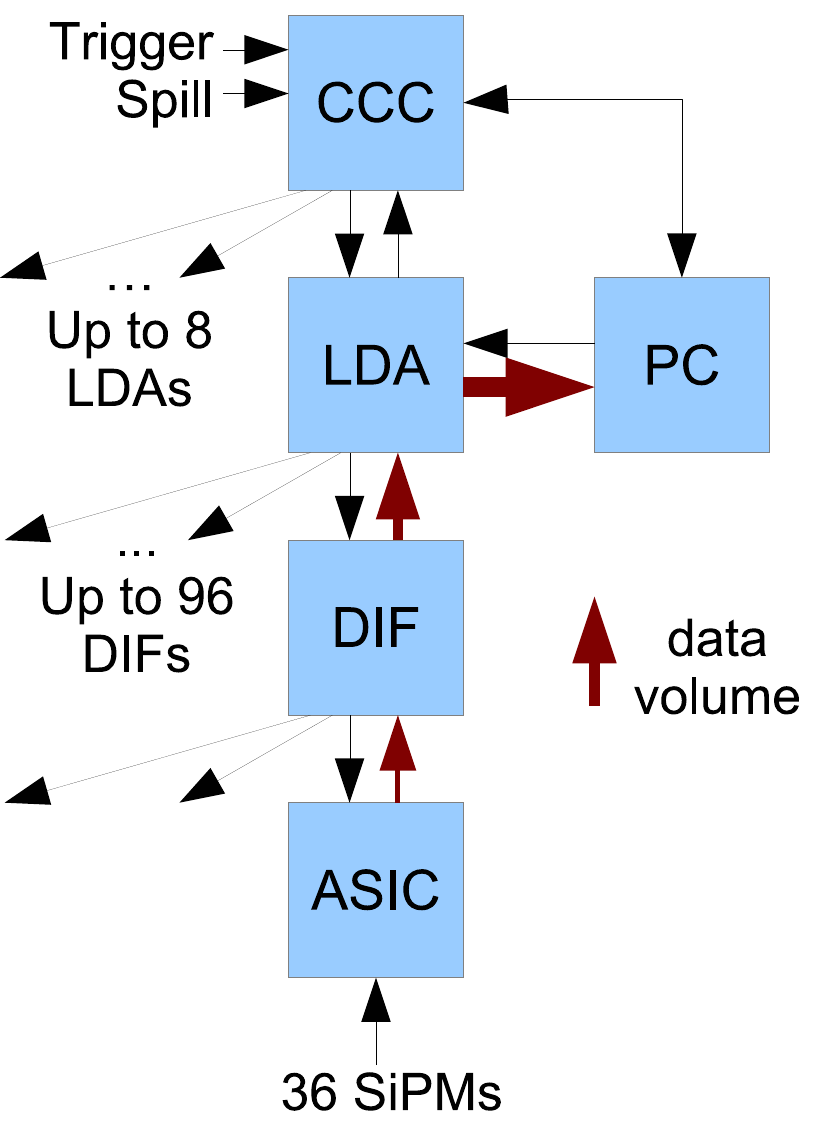}
     \caption{}\label{fig:DAQscheme}
  \end{subfigure}%
  \hfill
  \begin{subfigure}[T]{0.65\textwidth}
     \includegraphics[trim=0 0 0 20,clip,width=\linewidth]{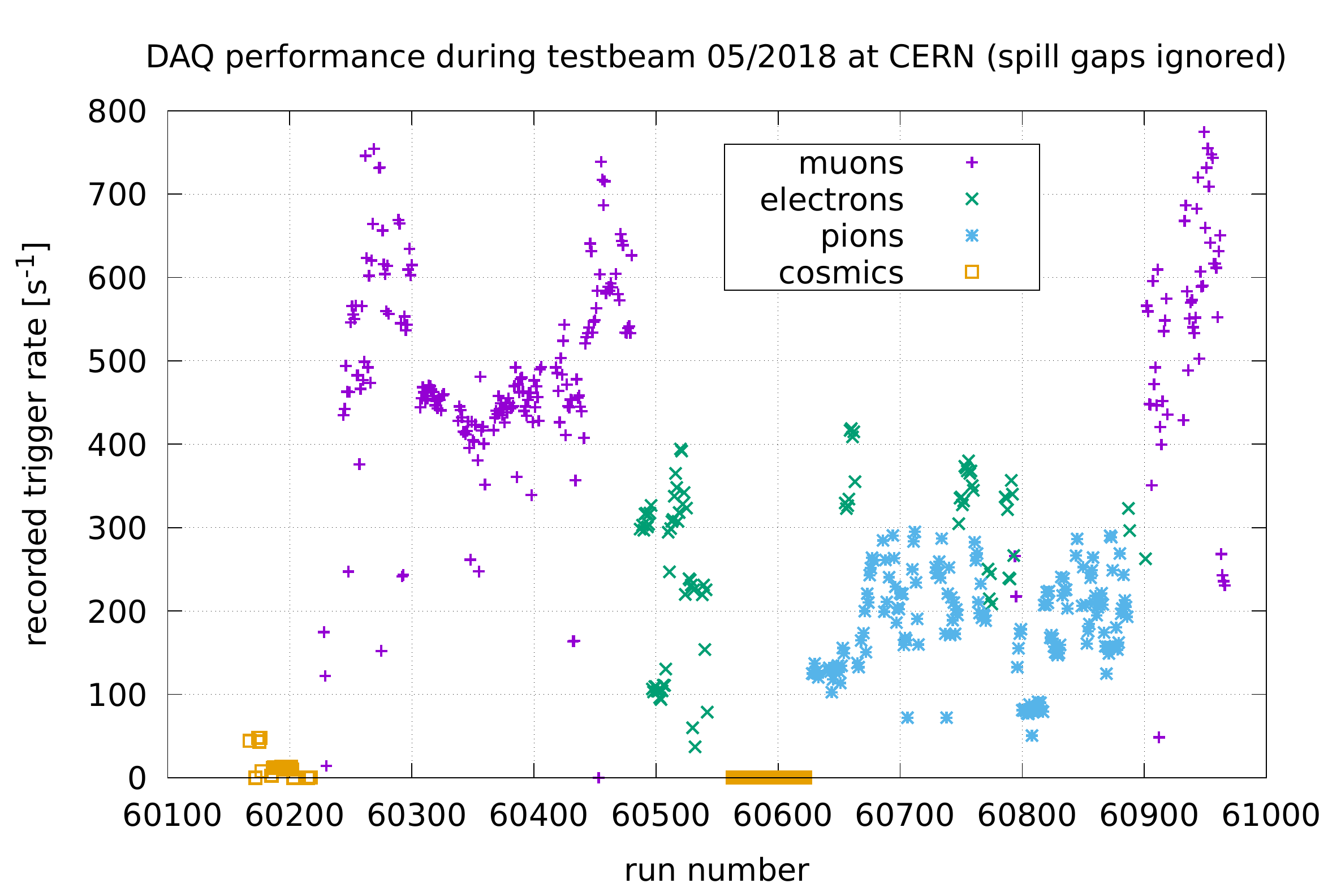}
     \caption{}\label{fig:DAQRates}
  \end{subfigure}%
  \caption{(\subref{fig:DAQscheme}) Sketch of the hierarchical structure of the DAQ system. (\subref{fig:DAQRates}) Example of sustained event rates obtained for different particle types during the May 2018 beam test.}
    \label{fig:DAQ}
\end{figure}

The readout electronics of the AHCAL technological prototype is based on common CALICE developments~\cite{ref:CALICEDAQ}. It follows a hierarchical structure as illustrated in figure~\ref{fig:DAQscheme}. 

%A common $40\,{\rm MHz}$ clock ensuring the synchronisation of all detector parts is provided by the Clock- and Control Card (CCC). This is connected to several Link Data Aggregators (LDAs), which distribute the clock as well as commands received from the CCC and a PC running the DAQ software to the detector layers. The LDAs also collect the data from the detector layers, aggregate them and send them to the PC. Each detector layer is equipped with a set of interface boards: the Detector InterFace (DIF) handling the communication between the LDA and the readout ASICs, the POWER board handling the power supplies for the complete layer, and the CALIB board handling the LED calibration system. These three interface boards are realised as mezzanines carried by the Central Interface Board (CIB). Signals from external devices like trigger scintillators are integrated into the AHCAL DAQ via the BeamInterFace (BIF).
Up to $36$ SiPM channels are read out by one SPIROC2E ASIC. The communication of all ASICs within a layer is handled by the Detector InterFace (DIF) board, while the POWER board handles the power supplies for the complete layer, and the calibration (CALIB) board handles the LED calibration system. These three boards are realised as mezzanines carried by the Central Interface Board (CIB). The data of up to $96$ layers is aggregated by Link Data Aggregators (LDAs) and sent to a PC running the DAQ software. The LDAs also receive commands from the PC as well as from the Clock and Control Card (CCC) and distribute them to the DIFs. A common $40\,{\rm MHz}$ clock ensuring the synchronisation of all detector parts is provided by the CCC.

Even though the readout ASICs are integrated in the detector volume, no active cooling is foreseen there, in order to avoid dead regions in the detector and keep the material as homogeneous as possible. Therefore, the power consumption is strongly limited, with a design goal of $40\,{\rm \mu W}$ per channel, of which $25\,{\rm \mu W}$ are allocated to the readout ASIC. This goal can only be reached by adapting the readout electronics to the low duty cycle of a linear electron-positron collider. In their beam structure, short bunch trains ($1\,{\rm ms}$ long for ILC) are interleaved with long gaps without particles ($199\,{\rm ms}$ long for ILC). This is taken into account in the readout electronics by having two phases: a data acquisition phase in which the ASICs are acquiring and storing the data, and a readout phase in which the data are digitised and sent from the ASICs via the LDAs to the PC. During the readout phase, all non-necessary components of the ASIC can be switched off {\it (power pulsing)}. The synchronisation of the phases between all ASICs can be ensured by a SPILL signal centrally provided by the CCC, or by a BUSY signal generated by the ASICs. 

\subsubsection{SPIROC2E Readout ASIC}
The SPIROC2E~\cite{ref:SPIROC} is a dedicated very front-end ASIC for SiPMs in calorimetric measurements, developed by the OMEGA group, realised in AMS SiGe $0.35 {\rm \mu m}$ technology. It features a large dynamic range, low noise, low power consumption and high precision for the amplitude measurement. It has $36$ channels with variable gains to achieve charge and time measurements. Each channel embeds an input DAC to tune the SiPM bias voltage, by subtracting the channel-wise adjustable value from the globally supplied bias voltage. The tuning range can be chosen in slow control settings to reach either up to $2.5\,{\rm V}$ or up to $4.5\,{\rm V}$ The ASIC is optimised for charge measurements in the range of $1$ up to $\sim\!3,000$ photo-electrons, corresponding to $160\,{\rm fC}$ up to $480\,{\rm pC}$ at a SiPM gain of about $10^6$. For each of the $36$ channels, two input preamplifiers, for a high gain and a low gain measurement, are needed to handle the required large dynamic range. Both are followed by variable slow shapers, for which a shaping time of $50\,{\rm ns}$ was chosen. The high gain preamplifier is in addition connected to a trigger line with a fast shaper and a discriminator, providing channel-wise self-triggering for thresholds as low as $50\,{\rm fC}$ ($1/3$ of a photo-electron). If one channel triggers, the data for all channels of this ASIC are kept.

The time measurement is performed with a coarse $16$-bit counter related to the bunch crossing clock and a fine time ramp based on this clock. The ramp has alternating directions from one bunch crossing to the next to avoid dead time connected with a ramp reset. The bunch crossing clock can be chosen between $250\,{\rm kHz}$ and $5\,{\rm MHz}$. The time resolution depends on the bunch clock, and is expected to reach $1\,{\rm ns}$ with a $5\,{\rm MHz}$ bunch clock. For low data rates also the data taking efficiency depends on the bunch clock, since the data acquisition should be stopped and the detector be read out when the bunch crossing counter overruns. In order to optimise data taking efficiency, the AHCAL prototype has been operated with a $250\,{\rm kHz}$ bunch clock during the beam tests so far, except for a few test runs.

An analog memory array with a depth of $16$ cells for each channel is used to store the time information and the charge measurement. If all $16$ {\it memory cells} are filled, the ASIC provides a BUSY signal via the LDA to the CCC. This is used to stop the data acquisition phase and switch to the readout phase for all channels of the detector. In addition, the length of the acquisition phase is limited by a time-out to avoid decay of the analog signals in the memory cells and an overrunning of the bunch clock counter. For AHCAL operation this time-out is typically set to $\sim\!15 {\rm ms}$.

In the readout phase, the analog memory content is digitised with an internal $12$-bit Wilkinson Analog-to-Digital Converter (ADC) and stored in a RAM. Out of the three measurements per channel (high gain, low gain and time), only two can be digitised and read out. The standard operation mode for beam tests was the {\it auto-gain} mode in which the ASIC provides the time measurement, and one of the two charge measurements, depending on the signal amplitude. Dedicated runs where both charge measurements (and no time information) have been read out are used to determine the inter-calibration between the two gain modes. 

In addition to the self-triggered mode, the SPIROC2E ASIC can also operate in a mode where the signal is sampled with a fixed delay with respect to an external trigger signal. This mode is used for the calibration with LED signals. 

The SPIROC2E ASIC is optimised for low power consumption. Its actual value depends on the time periods for which the ASIC components are switched on, including a short period before the expected arrival of events to allow the ASIC to stabilise. For optimised timing settings, the power consumption in ILC mode has been measured to be about $100\,{\rm \mu W}$~\cite{ref:SPIROC2EPower}, about a factor of four higher than the design goal. In future SPIROC versions, the power consumption can be reduced by digitizing and reading only the data of triggering channels.
%\todo{where does the actual power consumption fit? it's about 100uA/channel instead of goal of 25. cite Timo's thesis}

\subsubsection{Detector InterFace (DIF)}
The DIF~\cite{ref:AIDA2020_MS58} is based on a Zynq-7020~\cite{ref:Zynq7000} SystemOnChip (SoC). It can handle the communication of up to $72$ readout ASICs, the maximum foreseen in a layer in the ILD barrel, consisting of three parallel slabs of up to six HBUs. All ASICs in a slab are read out via two parallel readout chains, leading to up to six readout chains per DIF. The DIF generates the slower bunch clock from the central $40\,{\rm MHz}$ clock. The CIB, on which the DIF sits as a mezzanine, provides the connection to the central slab and an HDMI connector for the connection to the LDA. In addition, a single HBU can be operated without CCC and LDA via a direct USB connection from the CIB to the DAQ PC, allowing for easy tests in the lab. Additional slabs on the left and right can be connected with passive SIBs to the CIB.
 
\subsubsection{Link Data Aggregator (LDA)}
The LDA is also based on the Zynq 7020 SoC. It is available in two versions: the mini-LDA and the wing-LDA. The mini-LDA is realised as a dedicated mezzanine board for the ZedBoard, a commercially available development board for the Zynq. It has HDMI connectors for up to $10$ DIFs and has been used in many beam tests with smaller setups. The wing-LDA has been designed with the geometry of the ILD barrel sectors in mind. It has two times $48$ micro-HDMI connectors to handle two barrel sectors. It is realised as three passive boards, hosting the micro-HDMI connectors, connected to an active central board. The active central board has a slot for a Mars ZX3 module holding a Zynq 7020, and four slave Xilinx Kintex7 FPGAs for an efficient aggregation of the data volume. In the beam tests of the AHCAL technological prototype, the wing-LDA with only two passive boards and equipped with only two FPGAs has been used. In this configuration it can handle up to $48$ layers. Both LDA versions have an HDMI connector for the communication with a CCC.

\subsubsection{Clock and Control Card (CCC)} 
The CCC is based on the ZedBoard, which is extended with a dedicated mezzanine board with three HDMI connectors for the communication with LDAs and two LEMO connectors for externally provided SPILL and clock signals. In the beam tests, the CCC was used to generate the common $40\,{\rm MHz}$ clock and to distribute the clock and the SPILL signal to one or two LDAs. The operation of up to eight LDAs can be achieved with a fan-out.

\subsubsection{Beam InterFace (BIF)}
The BIF is based on the prototype of an AIDA mini-TLU~\cite{ref:miniTLU,ref:TLUpaper}, which provides a simple and flexible interface for fast timing and triggering signals. It receives the $40\,{\rm MHz}$ clock from the CCC on a LEMO connector, and the start and stop signals of the AHCAL data acquisition phases on a RJ45 connector. The timestamps of up to four external signals provided on LEMO connectors can be recorded with a precision of $\sim 1\,{\rm ns}$.

\subsection{LED Calibration System}
Due to the sensitivity of the SiPMs to changes of the operating temperature and the bias voltage, a gain calibration and monitoring system is necessary. The calibration system takes advantage of the property of the SiPMs to show a characteristic single pixel spectrum (SPS) at low light-intensities. This allows monitoring of the readout response to a well-defined signal, corresponding to one fired SiPM pixel, without the need to know the injected light intensity. This makes the calibration system also insensitive to possible changes of light intensity due to ageing of components. The HBUs house one LED per tile, producing the necessary light for calibration. The required fast driving pulses are transmitted electrically from the CALIB board to LED drivers on the HBU. When the calibration system is in idle-state, a capacitor is charged to the calibration bias voltage via a large resistor. Upon a LED trigger, a short pulse opens a transistor, and a current corresponding to the stored charge on the capacitor flows through the LED. This LED driver circuit guarantees a fast fall time of the optical pulse by a fast discharge of the LED's cathode node. The resulting optical pulse shapes have excellent small widths below $10\,{\rm ns}$~\cite{ref:LEDSystem}. The LED amplitude can be controlled by the calibration bias voltage. 

\subsection{Voltage Supply and Temperature Compensation}
\label{TemperatureCompensation}
The regulation of the low voltage for the ASIC power supply, the high voltage for the SiPM bias supply and the voltage for the LED calibration bias supply are performed on the POWER board. External supplies of $4.5\,{\rm V}$ (ASIC), $12\,{\rm V}$ (LED) and a bias voltage of at least $10\,{\rm V}$ above the desired operating voltage of the SiPMs are needed. In the beam tests of the AHCAL technological prototype, these have been provided by MPOD power supply modules~\cite{ref:MPOD}. The cables between the POWER boards and a cable distribution box were grouped by layer, while they were grouped by voltage between the cable distribution box and the MPODs. Here cables with a length of about $6\,{\rm m}$, corresponding to the expected length inside ILD, were used. 

The POWER board provides up to $3$ different SiPM bias voltages to a layer. They can be controlled by the LabView DAQ software within a range of $\sim 4\,{\rm V}$ with a precision of $\sim 40\,{\rm mV}$. In order to compensate for changes of the SiPM breakdown voltage caused by temperature variations, and thus keep their overvoltage constant, the SiPM bias voltage can be adjusted automatically according to the average temperature measured by up to six temperature sensors within a layer. This procedure has been set up with the measured mean dependence of the breakdown voltage on the temperature (section~\ref{sub:SiPMTest}). During the beam tests, it was set to run at every run start, and in fixed time intervals between ten minutes and one hour. 

\subsection{Online software} 
The dedicated DAQ and control software for the AHCAL technological prototype is based on LabView. It provides not only the fast commands for the data taking, but also the functionality for the configuration of the readout ASICs, the LDAs and the LED calibration system. This software has been interfaced to EUDAQ2~\cite{ref:EUDAQ2}, allowing easy integration with other devices (e.g. beam instrumentation such as wire chambers or beam telescopes) and user-friendly operation for non-experts.

In the EUDAQ2 {\it Producer} for the AHCAL technological prototype, several methods for event building have been implemented. Within the AHCAL, the synchronisation of the self-triggered hits is based on the {\it bunch crossing ID} (BXID), the cycle number of the slow bunch clock. Synchronisation with other detectors can be achieved with the help of a time stamp with $25\,{\rm ns}$ resolution or a trigger number. For this purpose, an external trigger signal, e.g. from a trigger scintillator, can be supplied to the AHCAL DAQ hardware.

The synchronisation with other detectors and the data storage was handled by EUDAQ2 {\it Data Collectors}. The data were stored in a raw data format, and as LCIO~\cite{ref:LCIO} files for analysis. To reduce the data volume, pure noise events were suppressed in the LCIO file by storing only data for BXIDs with an external trigger signal. However, in the raw files all data were kept, such that a later re-creation of the LCIO files was possible.

The readout speed depends strongly on the event size, which varies in self-triggered events due to the online zero-suppression. For the configuration with $38$ modules, the typical sustained event rate was between $120$ and $400$ electron or pion showers per second, reaching up to $600$ events per second for muons. An example of event rates obtained during the beam spills in the data taking in May 2018 is shown in figure~\ref{fig:DAQRates}. For LED calibration data, where all channels need to be read out, the online conversion to LCIO turned out to be a bottleneck. Therefore, these events were first stored only in raw data format and afterwards converted to LCIO. Nevertheless, the rate was limited to about $80$ LED triggers per second. Typical sizes for one event in raw (LCIO) data format were $7$--$10\,{\rm kB}$ ($3$--$5\,{\rm kB}$) for muons,  $10$--$17\,{\rm kB}$ ($3$--$11\,{\rm kB}$) for electrons and  $28\,{\rm kB}$ ($18\,{\rm kB}$) for $100\,{\rm GeV}$ pion showers.
%AHCAL_RAW:
%muon: 7 ~ 10 kBytes
%electron: 10 ~ 17 kBytes
%pion: 10 ~ 30 kBytes
%pion 100GeV: 28 kB
%SLCIO:
%muon: 3 ~ 5 kBytes
%electron: 3.5 ~ 11 kBytes
%pion: 5- ~ 23 kBytes
%pion 100GeV: 18 kB

During the beam tests, several methods for data monitoring were used. A very first cross check was provided by the LabView DAQ program, which shows event rates and raw data information per ASIC. EUDAQ2 also provides a simple online monitor, which was used to display hit positions in the layers and monitor the beam position. DQM4HEP~\cite{ref:DQM4HEP} was used to monitor event quantities, based on the directly written LCIO files. The most sophisticated monitoring was based on fully reconstructed data (with a preliminary calibration determined during the commissioning of the detector), with a dedicated program showing pre-defined histograms, and with an event display. This information could be accessed, based on partial files, already during the run, and was typically available within a few minutes after the run end for a complete run.  

\subsection{Absorber}
\label{sub:Absorber}

In a collider detector, the total depth of the calorimeter has to take into account the requirement to fit inside the diameter of the solenoid coil within the detector as well as the need to minimise longitudinal leakage. For ILD, the optimisation for jet energies up to $250\,{\rm GeV}$ leads to $48$ AHCAL layers in the barrel section, corresponding to $6$ interaction lengths for electromagnetic and hadronic calorimeters together~\cite{ref:ILCTDRDet}. 
The absorber of the AHCAL technological prototype is a self-supporting non-magnetic stainless steel structure with $44$ layers with a thickness of $16\,{\rm mm}$, corresponding to the inner part of the ILD barrel. 
Equipped with $38$ active layers, it covers $4.4$ nuclear interaction lengths. Therefore, at higher hadron energies, longitudinal leakage is expected to play a role, leading to a reduced reconstructed energy as well as a degradation in energy resolution, if no tail catcher is present to recover the leaked part of the shower.  

The absorber structure has been produced from standard rolled plates, which had undergone a cost-effective roller-levelling procedure, ensuring a flatness of $\pm 1\,{\rm mm}$ demonstrated over an area of $2 \times 2\,{\rm m}^2$. It is supported only by $5\,{\rm mm}$ thick side panels without additional spacers, such that the gaps can be filled completely with active elements. The absorber structure was produced as a mechanical test structure~\cite{ref:EUDETAbsorber}.
In order to maximise the thickness of the calorimeter in terms of hadronic interaction length in the volume given by the solenoid coil, the thickness of the active layers is kept as small as possible. Their total thickness amounts to $6.4\,{\rm mm}$, including the steel cassette providing mechanical support with a wall thickness of $0.5\,{\rm mm}$ on both sides. In order to facilitate the installation also of slightly too thick active layers in the prototype, the gap size of the absorber structure has been chosen conservatively as $8.7\,{\rm mm}$. Since no problems with the thickness of the active layers and their installation in the absorber structure was observed, a more aggressive design could be chosen in future.

\subsection{Cooling}
The AHCAL technological prototype has no active cooling within the active layers, which is possible thanks to the very low power consumption of the SPIROC2E ASICs. The interface boards, however, need cooling. For this, in ILD a leakless cooling system is foreseen, with a pipe mounted on the absorber on the back side of the interface boards (see figure~\ref{fig:XSecLayer}). 
%\todo{check if this is still true. K. Gadow? M. Reinecke?}
For the  AHCAL prototype, a more flexible solution has been chosen, based on over-pressure, and with separate cooling plates made from copper. They are mounted on the absorber, thermally decoupled by $\sim 1\,{\rm cm}$ thick plastic bars. The cooling plates are connected to the most relevant heat sources on the interface boards with thermally conducting pads. Flexible tubes connect the cooling plates to common inlet and outlet pipes. The dimensions of the pipes are chosen to resemble the ILD design. The pipes are then connected by hoses to an external cooling unit, which runs with nominal settings of $1\,{\rm bar}$ and $20\,^\circ{\rm C}$. The cooling system worked reliably during the beam tests, keeping the interface boards below $45\,^\circ{\rm C}$.

%% file: construction.tex
\section{Construction and Quality Assurance}
\label{sub:Construction}

The construction and quality assurance procedures for the AHCAL technological prototype have been developed for scalability and mass production. Many steps follow standard procedures available in industry. The discussion here covers mainly the steps that include procedures developed for this prototype: the quality assurance of SiPMs and ASICs as well as electrically assembled HBUs (before tile assembly), the wrapping of scintillator tiles into reflective foil including tests of the wrapped tiles, and the assembly of the wrapped tiles on the HBU PCBs. Similar to most particle physics detectors, the construction and quality assurance were distributed over several sites: the electronics assembly of the HBU PCBs took place at DESY, after SiPM tests at the University of Heidelberg and ASIC tests at the University of Wuppertal. The scintillator tiles have been produced in Vladimir by the industrial partner Uniplast in cooperation with LPI and MEPhI. The tile wrapping and testing were performed at the University of Hamburg, the tile assembly on the HBUs and their tests with cosmic rays at the University of Mainz. The modules were assembled, commissioned and integrated into the prototype (see section~\ref{sub:commissioning}) at DESY, before the prototype was shipped to CERN for the beam tests. MPP Munich provided strong support for the calibration and analysis software.

All the HBUs with $3 \times 3\,{\rm cm}^2$ scintillator tiles were constructed according to the described procedures. The four HBUs with $6 \times 6\,{\rm cm}^2$ tiles were assembled manually at the University of Tokyo.

\subsection{Electronics Assembly}
All the electronics components on the HBU PCBs are assembled with standard pick-and-place and reflow soldering techniques. In a first step, the SiPMs are mounted and soldered on the bottom side. Then all the other components including the ASICs are mounted and soldered on the other side. As detailed below, for the SiPMs sample checks have been performed for each batch. All the ASICs have been tested and characterised before assembly. After the electronics assembly, a functional test of the HBU is done with the help of the LED system, which allows for a direct verification of each channel's signal quality from the observation of Single Pixel Spectra (SPS). The electronics assembly was done for a total of $160$ HBUs. 

\subsubsection{SiPM tests}
\label{sub:SiPMTest}
Since many commercial SiPMs have good quality, and the testing of the SiPMs poses a risk of damaging them, we decided to test only small samples of the SiPMs, which were then not used in the detector. Of each batch of $600$ SiPMs with the same bias voltage within $\pm 100\,{\rm mV}$, $576$ devices are needed for a module. These batches were delivered on reels for the pick-and-place machine, which discards a few devices at the beginning of the reel. We therefore decided to test $16$ SiPMs for each of the $40$ batches. For technical reasons, in two of the batches, only $12$ and $15$ devices were tested.

The setup used for the SiPM test was originally designed for tests of scintillator tiles with a SiPM coupled to their side, and has been modified for the measurement of surface-mount SiPMs~\cite{ref:AIDA2020_MS13}. To achieve a large through-put, the system relies on the parallelised measurement of $12$ SiPMs using a pulsed laser source, which allows acquiring sufficient statistics for the characterisation and the measurement of parameter spreads at a rate of $\sim\,10$ seconds per SiPM. It consists of a laser head with $12$ fibres, and a base plate that can hold up to $144$ SiPMs. The SiPMs are read out with $12$ KLauS2~\cite{ref:Klaus2} ASICs, whose signals are multiplexed and sent to a 12-channel ADC. The tests were done in a temperature-controlled environment at $25\,^\circ{\rm C}$. 

\begin{figure}[htb]
  \begin{subfigure}[T]{0.44\textwidth}
     \includegraphics[width=\linewidth]{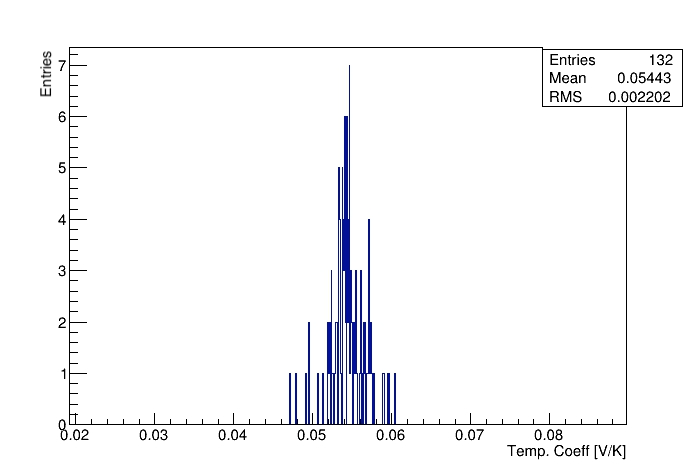}
     \caption{}\label{fig:dVdT}
  \end{subfigure}%
  \hfill
  \begin{subfigure}[T]{0.52\textwidth}
     \includegraphics[width=\linewidth]{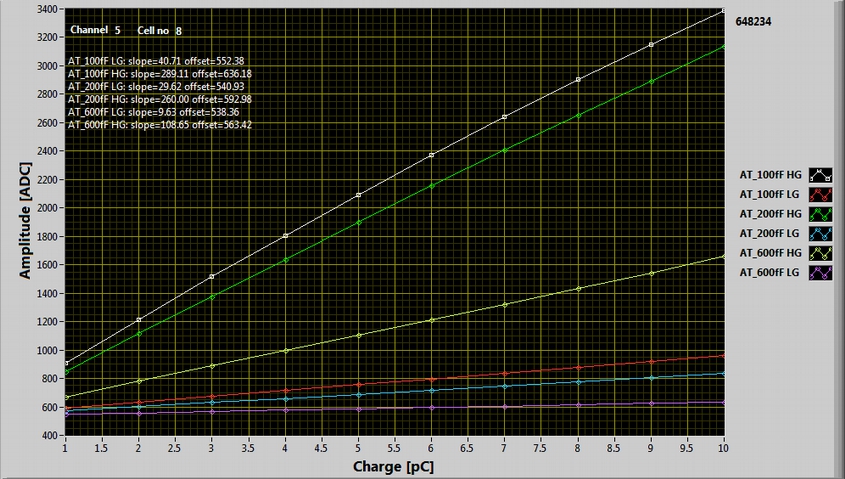}
     \caption{}\label{fig:ADCLinearity}
  \end{subfigure}%
  \caption{(\subref{fig:dVdT}) Distribution of the temperature coefficient for breakdown voltages, measured for a sub-sample of the SiPMs during the quality assurance tests. (\subref{fig:ADCLinearity}) Examples of the dependence of the amplitude on injected charge measured in self-triggered mode during the ASICs tests; the colours correspond to the three settings of the preamplifier feedback capacitors in high gain and low gain mode used for the characterization of the ASICs.}
    \label{fig:QATests}
\end{figure}

The measurements~\cite{ref:AIDA2020_D14.2} comprise the breakdown voltage, the dark count rate, the cross talk and the gain for all devices. The temperature dependence of the breakdown voltage was determined only for a sub-sample, since it is a very time consuming measurement.
Out of the $40$ batches, $5$ showed one outlier outside of the required $200\,{\rm mV}$ window for the breakdown voltage. The maximum window size containing all breakdown voltages of a batch was $270\,{\rm mV}$. At $5\,{\rm V}$ overvoltage, the average dark count rate was measured to be $74\,{\rm kHz}$, in agreement with the data sheet and well below our requirement of less than $500\,{\rm kHz}$. The cross talk was on average $4\%$, but due to a rather long integration window this number might also contain after-pulses. The average gain was determined as $6.2 \times 10^5$, with a device to device spread of $2.4\%$. Within a batch, the average gain spread was $1.8\%$. The average temperature dependence of the breakdown voltage was found to be $54.4\,{\rm mV/K}$ (see figure~\ref{fig:dVdT}), leading to an expected gain change of about $1\,{\rm \%/K}$ at $5\,{\rm V}$ overvoltage. The spread of the temperature dependence is small ($2.2\,{\rm mV/K}$) which is crucial for a common temperature compensation with the SiPM bias voltage of all channels in a layer. 
Due to the low dark count rate we decided to tolerate the cross talk rate, which is slightly higher than our original requirement of less than $3\%$. All batches passed the acceptance criteria.

\subsubsection{ASIC tests}
From previous productions, a yield of $80-90\%$ was expected for the ASICs. Since they are central for a fully functional and reliable detector, all ASICs were tested and characterised before assembly.  In order to test the SPIROC2E ASIC in an environment as similar as possible to the HBU, a test board has been designed that closely resembles a quarter of an HBU. The ASIC under test is inserted into a dedicated test socket for the BGA-372 package of the SPIROC2E. The test board provides a connector for the measurement of the input DACs. The input of an external pulse generator can be injected via a LEMO connector, and distributed to all $36$ channels of the ASIC via amplifiers that can be individually enabled or disabled by jumpers. Since the ASIC is very sensitive to the shape of the input pulse, the test board is also equipped with $8$ SiPMs that can be used to provide a real SiPM signal. For ASIC configuration and readout, the test board is connected to a set of interface boards (CIB, DIF, POWER, CALIB), which are steered from a mini-LDA and a CCC.
The measurement of the input DACs was performed with an Embedded Local Monitor Board (ELMB), which has originally been designed for the ATLAS experiment. For the tests, a dedicated LabView program was used that was derived from the AHCAL DAQ software.

Around $850$ ASICs have been tested. Most measurements were performed for three settings of the preamplifier feedback capacitors of $100\,{\rm fF}$, $200\,{\rm fF}$ and $600\,{\rm fF}$, the setting foreseen for beam operation. The following measurements were performed:
\begin{description} 
\itemsep0em
\parsep0em

\item Input DACs: The linearity of the 8-bit input DACs for the channel-wise adjustment of the SiPM bias voltage was tested for six values in the full range between $0$ and $255$.

\item Hold Scan: For a fixed injected charge of $3\,{\rm pC}$, the optimal hold times for the three preamplifier settings were determined both in external and in self-triggered modes. 

\item Amplitude Linearity: The amplitude linearity was measured for the three preamplifier settings. In external trigger mode, $9$ injected charges between $1\,{\rm pC}$ and $17\,{\rm pC}$ were used, while in self-triggered mode, $15$ injected charges between $1\,{\rm pC}$ and $15\,{\rm pC}$ were used. An example is shown in figure~\ref{fig:ADCLinearity}.

\item TDC Ramp: The full range of the rising and the falling ramp of the TDC was explored by measuring a fixed injected charge of $1\,{\rm pC}$ for twelve delays between $0\,{\rm \mu s}$ and $8\,{\rm \mu s}$. This measurement was done only for a preamplifier capacitance of $200\,{\rm fF}$.
\end{description}

The time needed to perform all tests for one ASIC was about $10$ minutes.
Only ASICs with all $36$ channels working were accepted for assembly. The resulting yield was $89\%$, in agreement with expectations for such a complex ASIC with analog and digital elements.

\subsubsection{HBU Functional Test with LEDs}
Once the electronics assembly is done, a functional test of the HBU can be done with the help of the LED system. For this the HBU is placed with the SiPMs facing a reflective surface and connected to the standard DAQ chain. This tests the complete signal chain, but does not yet allow a calibration, since the light output is not defined without the tiles. The test has been performed for all HBUs, and a few problems were spotted. 
%Out of three sets of SPIROC2E, one set had not been kept in a low oxygen atmosphere all the time before soldering, leading to bad soldering. All the affected $64$ ASICs could be replaced, and the HBUs were fully functional afterwards. 
%For the other SPIROC2E sets, a relatively high fraction of about $20\%$ of badly soldered ASICs was observed in the beginning, probably caused by non-optimal reflow soldering parameters. 
%In the beginning, a relatively high fraction of about $20\%$ of badly soldered ASICs was observed, probably caused by non-optimal reflow soldering parameters. This was repaired by a second reflow soldering for the affected boards, and avoided by a longer soldering for the remaining HBUs. The additional stress of the second soldering did not lead to any further effects. 
Two bad HBUs were identified which have a short within the PCB that was impossible to repair. This led to only $38$ modules being assembled in the end instead of the $40$ we had originally planned. On the $158$ working HBUs, in total one dead channel (out of $22,752$ channels) was found. 

\subsection{Tile Wrapping}
\label{sub:Wrapping}
Individual wrapping of the tiles provides the best optical decoupling between neighbouring scintillator tiles, but requires a (semi-)automatic procedure for the cutting of the reflector and the wrapping due to the large number of tiles. We chose ESR foil produced by 3M reflector because it leads to a high light output of the tiles, but the high reflectivity is challenging for many optical instruments.

The cutting of the reflective foil was performed with a laser cutter. In addition to the cutting of the foil outline, the folding lines were "scratched"  on the outside with reduced laser power. This simplifies the folding during the wrapping procedure, but the depth of the scratch is very sensitive to environmental conditions. The cutting took about $9$ minutes per sheet of $25$ cut foil pieces.

For the wrapping, a dedicated machine was developed, with a carousel for foil supply, stepper motors for the wrapping process, and vacuum cups for the transport of the components.  In order to be accepted by the pick-and-place machine, the tiles including wrapping must have a well-defined maximum size with tolerance below $100\,{\rm \mu m}$. The wrapping machine has to ensure that the tightness of the wrapping foil is compatible with this requirement. A speed of slightly less than a minute per tile was reached.
 
\subsubsection{Tile Tests}
Since the tile wrapping process does not work if the tiles are not within tight mechanical tolerances of $\pm 50\,{\rm \mu m}$, the size was measured with a caliper for a sub-sample of the tiles before the wrapping. In the later assembly of the tiles, the size is checked optically by the pick-and-place machine. Since a bulging of the foil can influence the visible size of the wrapped tile, the QA check of the wrapped tile was not done with a caliper, but also optically. The size was determined with a commercial document scanner, using half an integrating sphere to ensure uniform illumination. The outline of the tile in the scanned image was found with the minimum bounding rectangle search in the ImageJ Shape Filter Plugin. In addition, spot checks of the light output of the wrapped tiles were performed with a $^{90}$Sr source. For a small sub-sample of the production tiles, detailed studies of the response uniformity were performed.

\subsection{Tile Assembly}
The wrapped tiles were glued to the HBU PCBs with Araldite 2011 two-component epoxy adhesive. The number, position and size of the glue dots was optimised, resulting in $8$ glue points per tile. The glue was applied to the PCB with the help of a screen printer and a pump printing stencil which protects the SiPMs. The distribution of the glue needed only a few minutes per HBU, but the cleaning of the screen took a considerable amount of time (more than an hour).

The tiles were placed on the HBU PCBs with a commercial pick-and-place machine. Most of the wrapped tiles were fed to the machine on reels. Special tiles with cut-outs for mechanical cassette fixtures ($6$ per HBU) were supplied on trays. The machine was operated in the slowest mode, yielding the highest precision. This resulted in an assembly time of $7$ minutes per HBU.
Within the pot life of the glue of about $100$ minutes, the tiles on $4$ HBUs, corresponding to one module, were assembled. The glue was left for curing over night. Afterwards, the HBUs were installed in a cosmic ray test stand, and operated for about $24$ hours. 

\subsubsection{HBU Test with Cosmic Rays}
\label{sub:HBUTest}
Since the HBUs are fully assembled at this stage, and the system of scintillator tile, reflective foil wrapping and SiPM is complete, the detector response to MIPs can be determined in the cosmic ray test stand. In addition a gain measurement can be performed with the LED system, allowing a first determination of the {\em light yield}, the MIP response in units of SiPM pixels (see also section~\ref{sub:Calibration}). 

A dedicated test stand for the tests of HBUs with assembled tiles with cosmic rays has been built~\cite{ref:AIDA2020_MS13}. The test stand has two trigger layers with an area of $365.0 \times  361.8\,{\rm mm}^2$ of scintillator strips read out by classical photomultipliers, rotated by $90^\circ$ to each other. They provide a position resolution of the size of an AHCAL tile. The total height between the two trigger layers is adjustable up to $50\,{\rm cm}$, allowing several HBUs to  be stacked between them and to be read out in parallel.  The setup is located in a dark box with temperature monitoring.
 
In total, $155$ HBUs were measured in the test stand, most of them with $4$ boards operated in parallel. In addition to the two HBUs already rejected in the functional tests with LEDs, three boards were not measured due to smaller effects that could be repaired later. For the first four tested HBUs, a mistake in the electronics assembly was found and corrected.

The $155$ HBUs were characterised in terms of gain and light yield for each channel. The overall gain and light yield were found to be very homogeneous: the gain spread within an ASIC was $1.7\%$. The light yield mean value was $14$ pixels per MIP with an RMS of $12\%$. The light yield measured for each cell was used as an initial calibration. Only $13$ of the $22,300$ tested channels showed an anomalous behaviour (noisy, dead or low light yield), corresponding to about $0.05\%$ of the total number of channels.

%% file: commissioning.tex
\section{Assembly and Commissioning}
\label{sub:commissioning}
The assembly and commissioning of the AHCAL technological prototype happened in steps: first, the modules were assembled, consisting of four HBUs and their interface boards in their cassettes. The modules are fully functional detector units, and they were tested individually with LED light. They were also tested in batches with cosmic rays and in beam tests at DESY. In this step, their final operating parameters were fixed. In a second step, the complete prototype including cooling was assembled, and exposed to cosmic rays.

\subsection{Module Assembly and Commissioning}
\begin{figure}[ht]
\centering
%  \begin{subfigure}[T]{0.4\textwidth}
\begin{tikzpicture}
%    \draw (0, 0) node[inner sep=0] {\includegraphics[angle=90,origin=c,trim=200 430 180 600, clip,width=0.5\linewidth]{./figures/Module.jpg}};
%    \draw (1, 1) node {Hello world};
    \node[anchor=south west,inner sep=0] (image) at (0,0) {\includegraphics[angle=90,origin=c,trim=200 430 180 600, clip, width=0.5\linewidth]{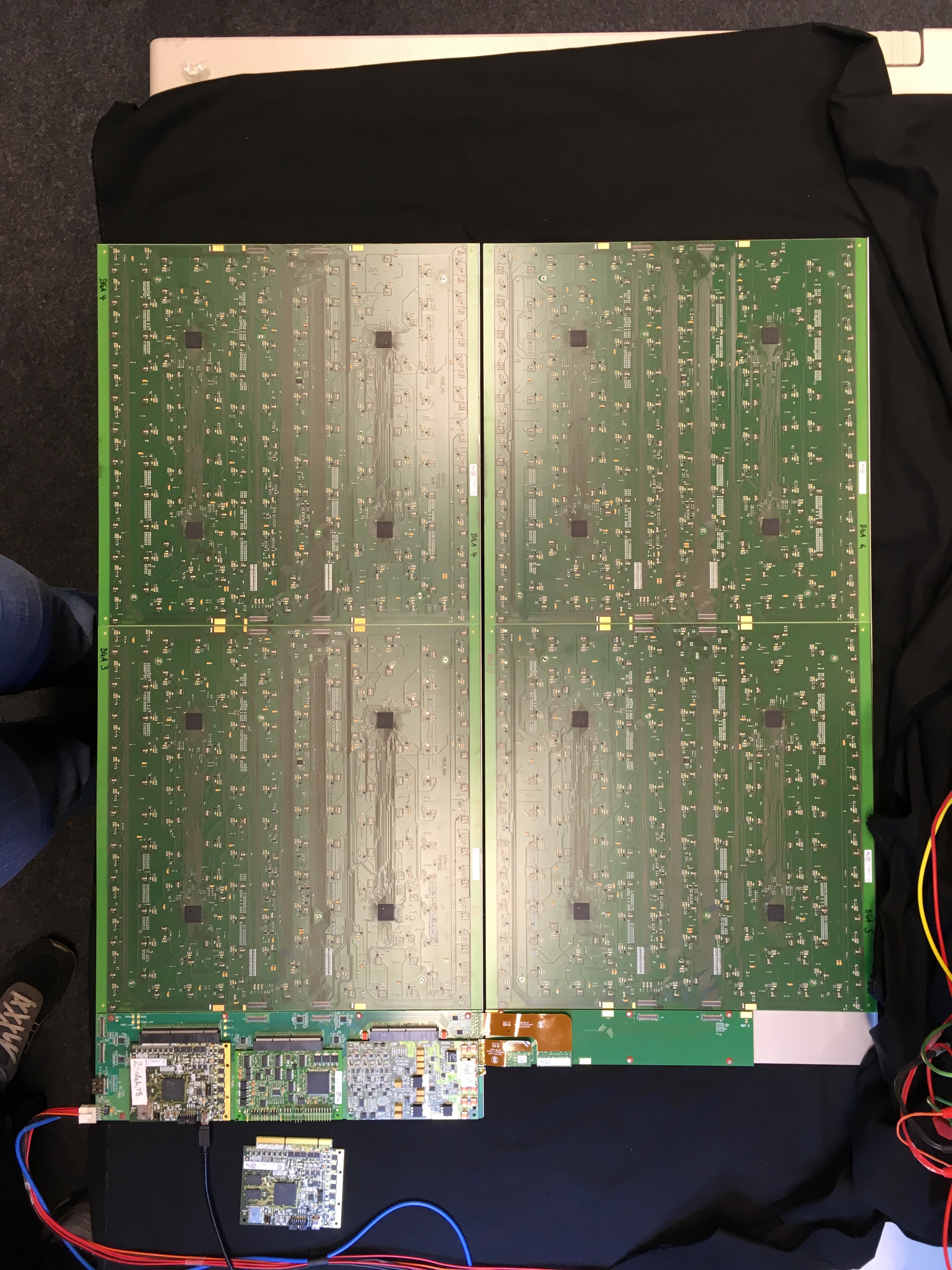}};
    \begin{scope}[x={(image.south east)},y={(image.north west)}]
        \draw[<->] (0.055,0.05) to  (0.87,0.05) ;
        \node[text width=2cm,fill=white] at (0.5,0.0) {$72\,{\rm cm}$};
    \end{scope}
%    \begin{scope}[x={(image.south east)},y={(image.north west)}]
%        \draw[help lines,xstep=.1,ystep=.1] (0,0) grid (1,1);
%        \foreach \x in {0,1,...,9} { \node [anchor=north] at (\x/10,0) {0.\x}; }
%        \foreach \y in {0,1,...,9} { \node [anchor=east] at (0,\y/10) {0.\y}; }
%    \end{scope}
%  \includegraphics[angle=90,origin=c,trim=200 430 180 600, clip,width=0.5\linewidth]{./figures/Module.jpg}
%     \caption{}\label{fig:Tile}
%  \end{subfigure}%
%  \hfill
\end{tikzpicture}
  \caption{Photo of one AHCAL module consisting of $2 \times 2$ inter-connected HBUs and the front end electronics (right side) comprising from top to bottom a SIB and a CIB with POWER, CALIB and DIF mezzanines.}
    \label{fig:Module}
\end{figure}

Four HBUs with SiPMs with the same operating voltage, a CIB with DIF, POWER and CALIB mezzanines and a SIB were assembled into a module (see figure~\ref{fig:Module}). These modules were kept in their respective configuration, no re-arrangement of components was done. The assembled modules were connected to a small DAQ system and tested with LEDs. Afterwards, they were installed in horizontal position in a dedicated test stand for cosmic rays~\cite{ref:Hodoscope}, that also provided position information with a few ${\rm mm}$ resolution, using a scintillator hodoscope constructed by the University of Tokyo. The last step in the module commissioning was a preliminary calibration performed in the DESY test beam facility with electrons of $3\,{\rm GeV}$. Without additional absorber plates, up to five modules (including cassettes) could be calibrated in parallel, with the electrons penetrating all modules as quasi-MIPs.
 
In the commissioning procedure, a number of operating parameters were fixed.
\begin{description}
\itemsep0pt
\parsep0pt
\item[SiPM bias voltage] First, the POWER board was configured for the SiPM operating voltage corresponding to an over-voltage of $5\,{\rm V}$ at $25\,{\rm ^\circ C}$. All SiPMs in a module were operated with the same voltage.  Therefore, the input DACs of the SPIROC2E ASICs were all set to the same value. A value of $50$ was chosen for the SPIROC2E input DACs,  corresponding to a voltage decrease of $3.7\,{\rm V}$. In this region the input DACs show a small channel-to-channel variation, and the value allows adjustments in positive and negative direction. In order to compensate for this decrease, the POWER boards were configured to provide a bias voltage of $8.7\,{\rm V}$ above the breakdown voltage of the SiPMs in the corresponding module.
\item[ASIC gain] The preamplifier capacitances were optimised for clearly visible Single Pixel Spectra (SPS) at low amplitudes and coverage of the full dynamic range of the SiPMs at high amplitudes. This led to $600\,{\rm fF}$ in the high gain branch and $1200\,{\rm fF}$ in the low gain branch for all modules, corresponding to a factor of about $20$ in gain between the two.
\item[hold time] The hold time was tuned with beam particles such that the pulse is sampled close to its maximum. 
%The value is the same for all channels on a chip.
Since the optimal hold time strongly depends on the signal shape, it can only be determined with signals from particles producing scintillation light. One parameter determines the hold time for all channels of the SPIROC2E. The same parameter value was found to be adequate for all ASICs.
\item[auto-gain threshold] The threshold for switching between the high gain and the low gain mode was chosen around $5$ MIP. This is well within the ADC range for the high gain mode, taking into account a non-linearity observed in the upper part of the ADC range in high gain mode, and a rather wide threshold behaviour of the gain switching, leading sometimes to high gain mode results for amplitudes well above the nominal auto-gain threshold. 
\item[trigger threshold] The trigger threshold was chosen as low as possible for a stable operation of many layers. Since the value is applied on the ADC scale, and the gain varies from channel to channel, the resulting threshold on the MIP scale also varies from channel to channel. The observed threshold values were around $0.25-0.3$ MIP.
\item[LED calibration voltages] The calibration voltage at which the LEDs start to emit light varies from channel to channel, both due to variations in the LED circuit and due to device to device variations. For an efficient cross check of the calibration, it is important to obtain analysable SPS for a large fraction of all channels with a small number of calibration voltages. It was found that three LED calibration voltages were sufficient to measure analysable SPS in about $95\%$ of all channels. These are used to monitor variations of the gain during data taking. The gain calibration used in the data analysis (see section~\ref{sub:GainCalibration}) is determined in extended calibration runs with more LED voltages. 
\end{description}

\subsection{Prototype Assembly and Test with Cosmic Rays}
After the successful commissioning of all modules, they were installed in the absorber stack. Based on the results from tests of the individual modules,  the few modules showing dead channels were placed in the back. The modules were installed and connected to the DAQ one by one, making sure at each step that installed layers were showing LED signals in external trigger mode. The main difficulty in this step are the flex-lead connections\footnote{For long-term operation in a collider detector, these connections would be sealed.} that sometimes mechanically disconnect and require a removal and re-installation of a module. 
%\todo{cannot read Felix' comment}

In this configuration, the LED calibration procedure was tuned to run with all layers in parallel. In addition, the full detector was operated in self-triggered mode for about $20$ hours. The detector was running stably during that time. Even though the active layers were in a vertical position and no further external trigger signal was available, events caused by cosmic rays could clearly be identified. This shows the very good quality of the detector in terms of noise and efficiency and the excellent uniformity of the SiPMs. 

%% file: calibration_split.tex
\section{Calibration}
\label{sub:Calibration}

The goal of the calibration procedure presented here is to convert the measurement from electronics units to a physics scale: for the amplitude measurement from ADC bins to the energy deposition of a MIP, and for the time measurement from TDC bins to nanoseconds. This equalises the response of all cells, and allows a prediction of the detector response to electromagnetic and hadronic showers. 

All measured energies are converted from the ADC scale to the MIP scale, according to
\begin{eqnarray}
E_i {\rm [MIP]} &=& \frac{(A_i {\rm [ADC]} - \mathit{Ped}_i {\rm [ADC]}) \times IC_i}{\mathit{MIP}_i} \times 
f_{{\rm sat},i}^{-1}(A_i {\rm [pixel]}) {\rm \ \ \ \ with}\\
A_i {\rm [pixel]} &=& \frac{(A_i {\rm [ADC]} - \mathit{Ped}_i {\rm [ADC]}) \times IC_i}{\mathit{Gain}_i}.
\end{eqnarray}
Here  $A_i {\rm [ADC]}$ is the measured amplitude in units of ADC bins for cell $i$, $\mathit{Ped}_i {\rm [ADC]}$ the corresponding pedestal, $IC_i$ the inter-calibration factor between the low gain and the high gain mode, and $\mathit{MIP}_i$ is a conversion factor of ADC bins into MIPs. The function $f_{\rm{sat},i}$ is the SiPM response function describing the SiPM output signal as a function of the incoming light intensity, normalised such that $f=1$ for small $A$. The inverse correction function depends on the number of fired pixels, $A_i {\rm [pixel]}$, calculated with the conversion factor of ADC bins into units of a single avalanche in a SiPM pixel, $\mathit{Gain}_i$.

The MIP conversion factor $\mathit{MIP}_i$ of each cell is determined using the approximately minimum ionizing muons, while the conversion factor to pixels $\mathit{Gain}_i$ is extracted from externally triggered LED runs. For each data taking mode, the corresponding pedestals are extracted from runs with the same conditions. The gain inter-calibration factor $IC_i$ is measured in runs in which the hit time information was dropped, and the amplitude was read out in high gain and in low gain mode {\it (inter-calibration mode)}. This is then used to convert low gain amplitude measurements into high gain units. All numbers in the following quoted in ADC bins refer to the high gain mode.

The values expected from the construction and commissioning are: about $15$ pixels fired for a MIP crossing a tile and a gain: $\mathit{Gain}_i \approx 16$ ADC bins per pixel, leading to a MIP conversion factor: $\mathit{MIP}_i \approx 240$ ADC bins.

In the following, each element of the amplitude calibration is discussed in further detail. At the end of the section, the calibration of the hit time measurement is described.

\subsection{Pedestal and Noise}
The pedestal value enters not only the measurement of every cell energy, but also the determination of the MIP conversion factor for each cell. Therefore, its accuracy impacts the precision of the energy scale of the detector, and has to be very small compared to the $\mathit{MIP}_i$ value.

The pedestal and noise values depend on the mode in which the SPIROC2E is operated, therefore they have to be determined in the mode corresponding to the data taking. Due to the analog storage of the data in {\it memory cells} before the digitization, the pedestal values depend not only on the channel, but also on the memory cell number. For the AHCAL prototype with $38$ layers, this leads to $38 \times 576 \times 16 = 350,208$ pedestal values per mode.

The default beam operation mode for the prototype is the self-triggered auto-gain mode. In this mode, the information for all channels is digitised and read out if at least one channel has a signal above the trigger threshold. In muon runs, where typically only one channel in an ASIC has a hit, the remaining channels can be used for the determination of the pedestals. To cover all memory cells, the beam rate has to be high enough to fill all $16$ slots. This method is used for the amplitudes measured with high gain in auto-gain mode.

The high gain pedestal values are typically around $530$ ADC bins, with a channel-to-channel spread (RMS) of about $33$ ADC bins. The spread of about $9$ ADC bins between memory cells of the same channel is significantly smaller, but not negligible compared to a MIP conversion factor of $\approx 240$ ADC bins. For memory cells $0$ to $8$, narrow pedestal distributions with a width of $3.5$ to $4$ ADC bins are observed, with very little indication of SiPM noise. For the higher memory cells, outliers are observed in the pedestal distribution, leading to large RMS values and shifted means. Therefore, we chose to apply a memory-cell-wise pedestal correction only for memory cells up to $8$, while we use the mean pedestal determined from memory cells $0$ to $8$ for memory cells $9$ to $15$ of the same channel. While the cause could not be identified yet, it has been checked carefully in the muon data as well as with charge injection on a test board that no such outliers are observed in the amplitudes of channels above the trigger threshold. 

It turned out that the method used for the high gain pedestals cannot be used to determine the pedestal for amplitudes measured with low gain since it leads to biased values. In data taken in inter-calibration mode, the most probable ADC value for muon signals was found to be lower than the ADC value in events without hits. Therefore, a method was developed to calculate the low gain pedestal from the muon signals measured with high gain and low gain in inter-calibration runs, using the inter-calibration factor determined from LED data. In addition, a correction for differences in the pedestal values observed for high gain between auto-gain and inter-calibration runs is applied. 
%Ped_{LG,AG} = (MIP_{LG,IC} - (MIP_{HG,IC}-Ped_{HG,IC})/IC) + (Ped_{HG,AG} - Ped_{HG,IC})
%MIP is amplitude in ADC, NOT pedestal subtracted 

\begin{figure}[htb]
\centering
%  \begin{subfigure}[T]{0.49\textwidth}
     \includegraphics[width=0.5\linewidth]{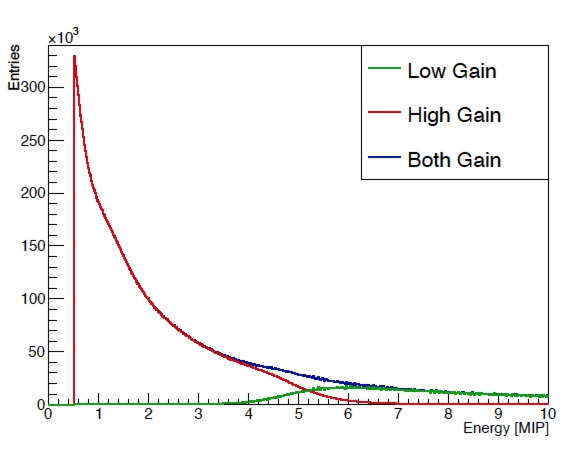}
%     \caption{}\label{fig:StackAtCERN}
%  \end{subfigure}%
%  \hfill
%  \begin{subfigure}[T]{0.49\textwidth}
%     \includegraphics[width=\linewidth]{./figures/LayersInStack.jpg}
%     \caption{}\label{fig:LayersInStack}
%  \end{subfigure}%
  \caption{Example of the low amplitude region of the reconstructed hit energy spectrum for all cells in a $100\,{\rm GeV}$ electron run taken in auto-gain mode. The amplitudes measured in high gain mode are indicated in red, while those measured in low gain mode are shown in green. The sum of both distributions (blue) shows a smooth transition.}
    \label{fig:HitEnergyHgLg}
\end{figure}

The low gain pedestals determined with this method are on average $16$ ADC bins lower than the value we would have obtained with the method used for the high gain pedestals. 
%This corresponds to a systematic shift towards lower amplitudes of about $1.4$ MIP, which would be introduced to all channels measured with low gain. It has been cross-checked with measurements for a few ASICs on a test board that the pedestal values determined with these methods for high gain as well as low gain are consistent with a linear extrapolation of the ADC values measured as a function of a charge injected into the ASIC channel. 
It has been cross checked with electron data that the used pedestal values lead to a smooth transition between the high gain and the low gain range in the hit energy spectra (see figure~\ref{fig:HitEnergyHgLg}).

The pedestals in external trigger mode were obtained in LED calibration runs with the LED voltage set to $0\,{\rm mV}$. No outliers were observed in these distributions, so individual values were determined and used for all memory cells. The high gain pedestal values in external trigger mode are typically around $500$ ADC bins, with a channel-to-channel spread of about $32$ ADC bins. The width of the pedestal distribution is around $4$ ADC bins, in agreement with the values obtained for self-triggered events. 
%The values are consistent between all three testbeam periods.

\subsection{Gain Calibration}
\label{sub:GainCalibration}
The conversion factor from ADC bins to pixels $\mathit{Gain}_i$ corresponds to the charge produced by a single avalanche in the SiPM, measured in ADC bins by the SPIROC2E readout ASIC, so it is the product of the SiPM gain (in electrons per avalanche) and the ASIC gain (in ADC per electron). It is therefore not directly comparable to the gain values quoted and measured for the SiPMs alone, but is an important quantity for the calibration and the monitoring of the detector in several ways: as described above, it is needed to convert the measured amplitude from ADC bins to pixels in order to apply the correction for SiPM non-linearity. Together with the MIP conversion factor $\mathit{MIP}_i$, it is used to determine the {\it light yield}, the MIP response in units of SiPM pixels. In addition, it can be used to easily monitor the  uniformity and stability of the detector, since it can be determined from short LED runs. While for the determination of the calibration constants long LED runs with many different LED voltages were used, the monitoring was performed with daily runs with only three LED voltages, measured in roughly half an hour. 
 
\begin{figure}[htb]
  \begin{subfigure}[T]{0.42\textwidth}
     \includegraphics[width=\linewidth]{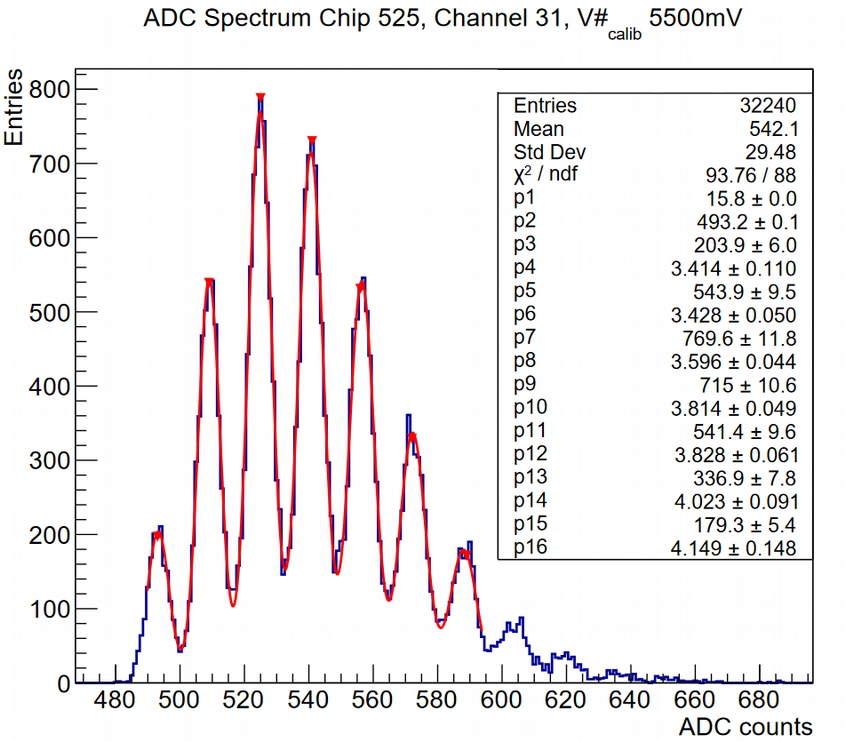}
     \caption{}\label{fig:SPS}
  \end{subfigure}%
  \hfill
  \begin{subfigure}[T]{0.55\textwidth}
     \includegraphics[trim=0 0 0 0,clip,width=\linewidth]{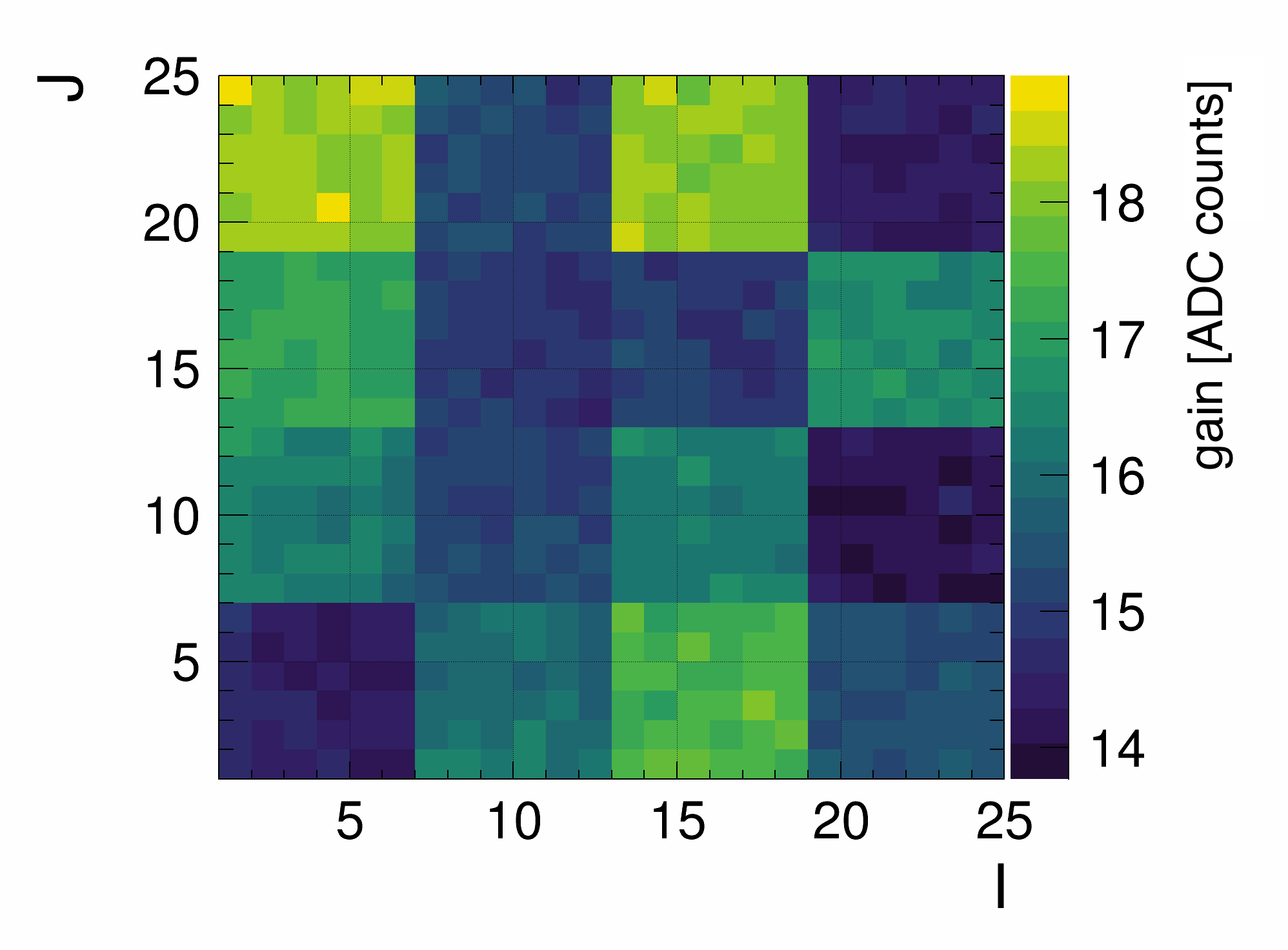}
     \caption{}\label{fig:GainLayer}
  \end{subfigure}%
  \hfill
  \caption{(\subref{fig:SPS}) Example of a single pixel spectrum with the fit with the sum of several Gaussian functions. ${\rm p1}$ and ${\rm p2}$ correspond to the gain and the position of the first peak, while the remaining fit parameters are the heights and widths of the individual peaks (see text). (\subref{fig:GainLayer}) Example of the distribution of the gain in a layer. I and J are integers representing the position of a channel in the layer.}
    \label{fig:Gain}
\end{figure}

The gain is determined from single pixel spectra obtained by illuminating the SiPMs with short pulses of LED light. An example spectrum is shown in figure~\ref{fig:SPS}, clearly showing the individual pixel peaks. It was obtained after correction of the memory-cell-wise shift of the pedestal. For each channel, the distribution is fitted with the sum of several Gaussian functions, in which the distance between adjacent peaks, corresponding to the gain, is constrained to be the same for all peaks, whereas the amplitude and the width of each peak is left as free parameter. . The mean gain is about $16$ ADC bins per pixel, with a channel-to-channel spread in the complete prototype of $1.1$ ADC bins per pixel. This spread has a large contribution from the gain spread between ASICs. Channels within an ASIC show on average a spread of only $0.24$ ADC bins per pixel ($1.5\%$). This is consistent with the value of $1.7\%$ measured on the individual HBUs (section~\ref{sub:HBUTest}) and average spread within a batch of $1.8\%$ measured in the SiPM sample tests (section~\ref{sub:SiPMTest}). The influence of the ASIC is also visible in the distribution of the gains within a layer (figure~\ref{fig:GainLayer}), with a clear pattern caused by the $16$ ASICs. In the long LED runs, the gain could be determined for typically $98\%$ of all channels. For the remaining channels, the average of the corresponding ASIC is used.

%The stability of the gain between the three beam test periods was excellent, with a mean shift of significantly less than a percent, and a channel-by-channel spread of the shift of around one percent. A larger shift observed for one module could be traced to a broken channel of the MPOD power supply. 

%\todo{should that go to testbeam section?}
%The daily LED runs were used to monitor the dependence of the gain on the temperature. Since the electronics within the absorber structure is not cooled directly, the average temperature of the prototype varies slowly with time. After a warm-up phase in the beginning, a day-night variation of the order of one degree is observed. The results for the temperature dependence~\cite{ref:TempComp} are shown in figure~\ref{fig:TempComp} for the beam period in October 2018. All gain values agree with each other to better than $1\%$, which is the expected accuracy due to the gain determination procedure and the granularity of the SiPM bias voltage adjustment.

%\begin{figure}[htb]
%  \begin{subfigure}[T]{0.49\textwidth}
%     \includegraphics[width=\linewidth]{./figures/GainComp_HVsetting.jpg}
%     \caption{}\label{fig:GainComp_HV}
%  \end{subfigure}%
%  \hfill
%  \begin{subfigure}[T]{0.49\textwidth}
%     \includegraphics[width=\linewidth]{./figures/GainComp_Gain.jpg}
%     \caption{}\label{fig:GainComp_Gain}
%  \end{subfigure}%
%  \caption{(\subref{fig:GainComp_HV}) Temperature and SiPM bias adjustment values and (\subref{fig:GainComp_Gain}) Temperature and Gain values during the October testbeam period. }
%    \label{fig:TempComp}
%\end{figure}

\subsection{Gain Inter-calibration}
Since the auto-gain mode is usually used in beam measurements, the inter-calibration factor $IC_i$ is needed to convert the energy of large amplitudes measured with low gain to the same scale as the amplitudes measured with high gain. In principle, both beam data and LED data taken in inter-calibration mode can be used to determine this factor. Due to the larger statistics and the difficulties to extract the low gain pedestal in self-triggered beam data, we used externally triggered LED data. The inter-calibration factor can be extracted in the amplitude range where both the high gain and the low gain are within the linear ADC range. In addition, one has to take into account that the SPIROC2E ASIC shows a shift of the pedestal if the total charge measured on the ASIC is large. This is the case for LED data at large calibration voltages, since many channels have a large signal then. We use two methods for the determination of the inter-calibration factor: (1) fitting the curve of the high gain response as a function of the low gain response in a large range of calibration voltages and (2) using the high gain and low gain amplitude differences between two LED runs with a small difference in calibration voltage. We prefer method (2) since it minimises the influence of the pedestal shift. For about 
$10,000$ channels in which both methods work well, on average a shift of $0.4\%$ is observed between the values obtained with the two methods, with a spread of $2\%$. For about $80 \%$ of all channels the inter-calibration factor could be determined with method (2), with a mean value of about $19.5$ and a channel-to-channel spread of around $4\%$. For the other channels this mean value is used. It is slightly lower than the value of $20$ expected from the ratio of the preamplifier capacitances. 
%The inter-calibration factor is stable between the beam periods, with a mean ratio of $1.00$ and a channel-to-channel spread of $2.3\%$.

\subsection{MIP Calibration}
\begin{figure}[htb]
\centering
%  \begin{subfigure}[T]{0.6\textwidth}
%     \includegraphics[width=0.6\linewidth]{./figures/MIPSpectrum.png}
     \includegraphics[width=0.6\linewidth]{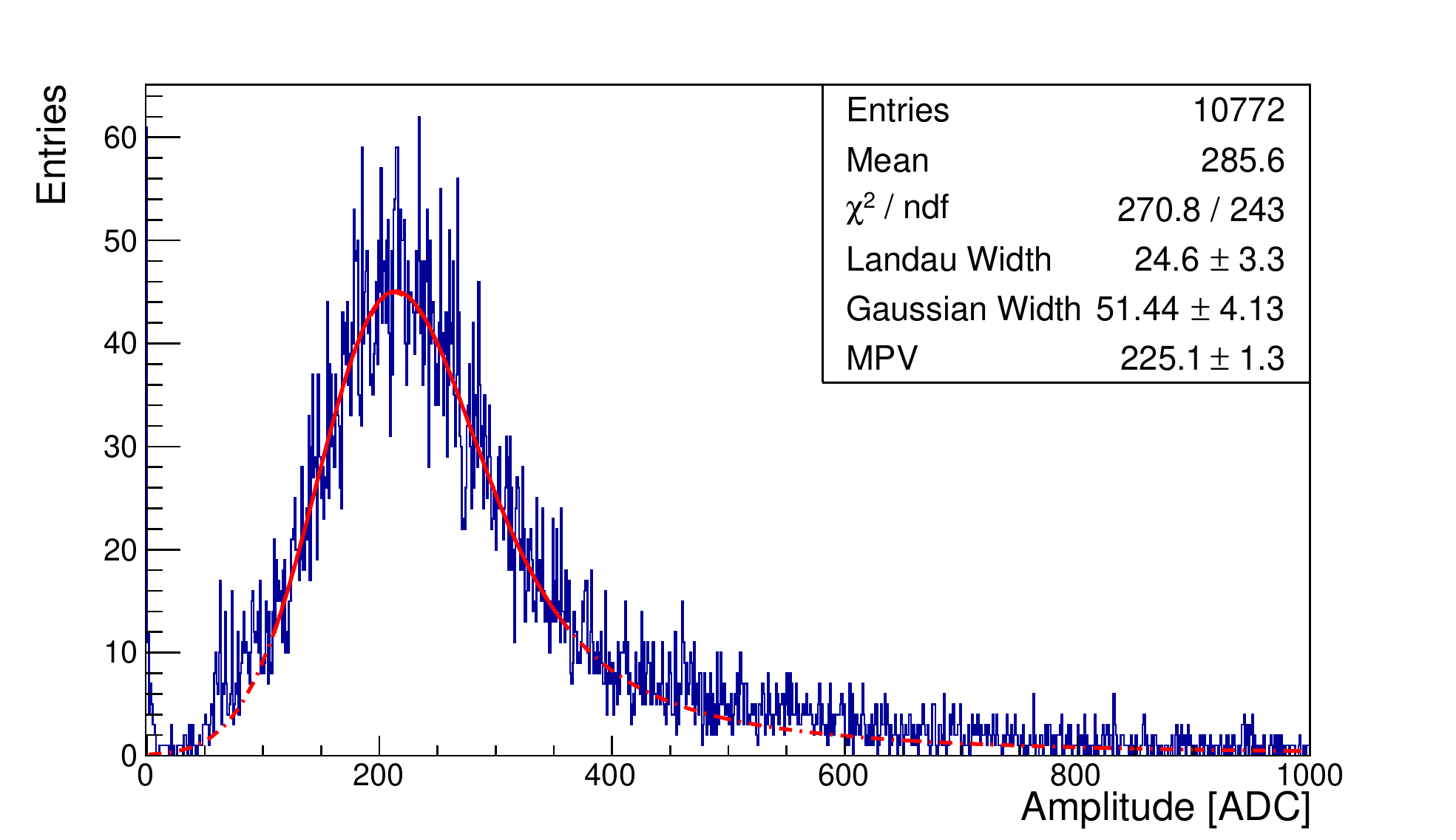}
%     \caption{}\label{fig:MIPSpectrum}
%  \end{subfigure}%
%  \hfill
%  \begin{subfigure}[T]{0.49\textwidth}
%     \includegraphics[width=\linewidth]{./figures/GainComp_Gain.jpg}
%     \caption{}\label{fig:GainComp_Gain}
%  \end{subfigure}%
  \caption{Exemplary MIP Spectrum for one channel after pedestal subtraction. The blue histogram shows the data, while the red line indicates the fit with the convolution of a Landau distribution and a Gaussian function (full line:\ fit range, dotted line:\ extrapolation). For a small number of events the ASIC triggers even though the hit amplitude is below the trigger threshold at $\approx\,60\,{\rm ADC\ counts}$.
  }
    \label{fig:MIP}
\end{figure}

The MIP conversion factors $\mathit{MIP}_i$ of each cell is extracted using the most probable response to the approximately minimum ionizing muons. A typical MIP spectrum is shown in figure~\ref{fig:MIP}. The calibration of each cell is accomplished with muons from a beam that has a broad spatial distribution in order to minimise the number of detector positions needed to cover the entire front face of the AHCAL. 
A minimum of $1,000$ muon events per channel is required to obtain a reliable fit to the pulse height spectrum. In practice, we aimed to have at least $10,000$ events per channel, since also the pedestals are extracted from the same data, for which a sufficient amount of events in the higher memory cells is needed. During the beam tests at the CERN Super Proton Synchrotron~(see section~\ref{sub:SPSTestbeam}), muon beams with energies of  $40\,{\rm GeV}$ and $120\,{\rm GeV}$ were used. For the $40\,{\rm GeV}$ muon beam, we scanned $17$ positions with $\approx\,500,000$ events per position, corresponding to about one hour per position (with $2$ spills of $4.5$ seconds length within a cycle of about $30$ seconds). For the narrower $120\,{\rm GeV}$ muon beam, we collected $\approx\,150,000$ events each at $49$ positions, leading to a similar amount of running time.

In the analysis of the muon data, no event or hit selection is applied except the coincidence with an external trigger scintillator already applied in the online event building. Due to the low SiPM noise and the clean muon beam, this results in clean MIP signals for nearly all channels. After a memory-cell wise pedestal subtraction, the resulting ADC spectra are fitted with the convolution of a Landau distribution and a Gaussian function. The broadening of the Landau distribution, which leads to the need of the convolution with a Gaussian, is to a large extent caused by the statistical fluctuations of the small number of detected photo-electrons for a MIP, which is about $14$ for the AHCAL. We define the value of the MIP conversion factors $\mathit{MIP}_i$ as the most probable value (MPV) of the convoluted function. The broadening leads to a shift of the MPV of the convolution with respect to the MPV of the Landau distribution. However, the fitted parameter values of the Landau and the Gaussian show large correlations, while the MPV of the convoluted function is more stable than the MPV of the Landau distribution.

For more than $99.9\%$ of all channels, good fits could be obtained. Of the remaining $19$ channels, $6$ were dead, while the other channels showed no clear MIP peak or the fit failed for other reasons. These channels are ignored later in the offline analysis.
The mean MIP conversion factor is $218$ ADC bins, with a channel-to-channel spread of $14\%$. 
%It is stable between running periods, showing on average a shift of $1\%$ with a spread of $2.7\%$ between May and June 2018.

\begin{figure}[htb]
  \begin{subfigure}[T]{0.42\textwidth}
     \includegraphics[width=\linewidth]{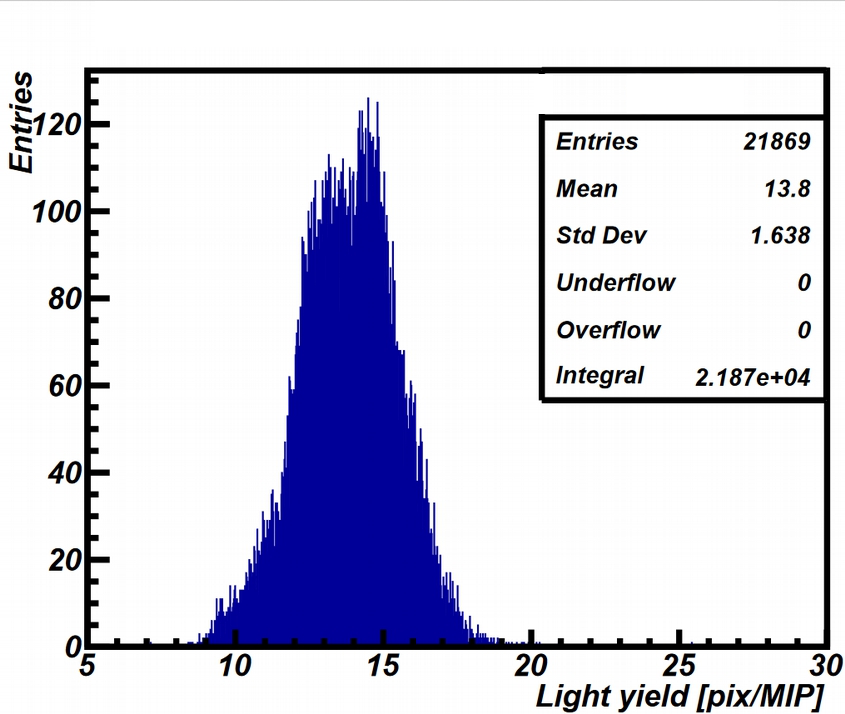}
     \caption{}\label{fig:LYHist}
  \end{subfigure}%
  \hfill
  \begin{subfigure}[T]{0.55\textwidth}
     \includegraphics[width=\linewidth]{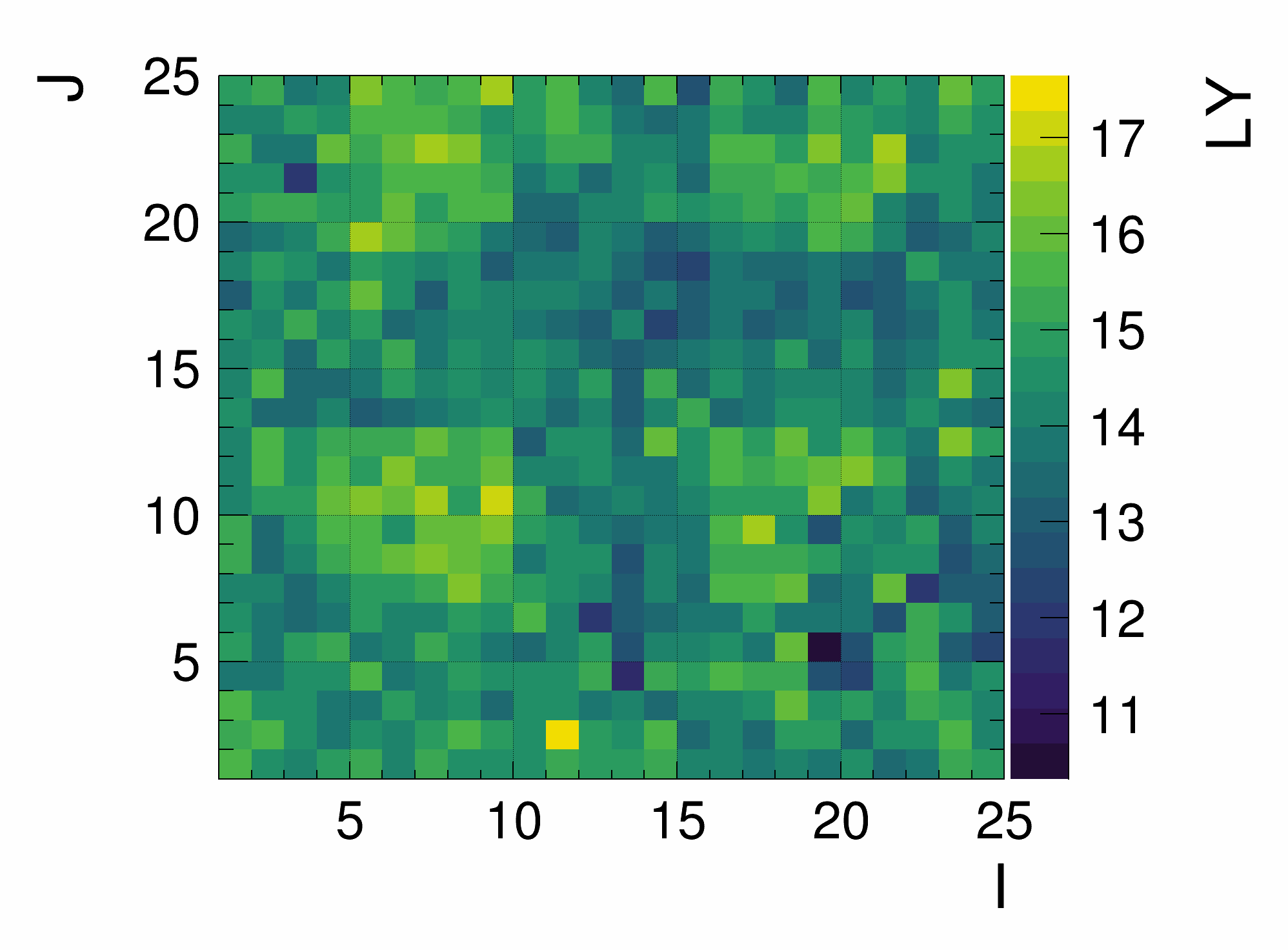}
     \caption{}\label{fig:LYLayer}
  \end{subfigure}%
  \caption{(\subref{fig:LYHist}) Light yield distribution. (\subref{fig:LYLayer}) Example of the distribution of the light yield in a layer. I and J are integers representing the position of a channel in the layer. }
    \label{fig:LY}
\end{figure}

From the MIP conversion factor and the gain, the light yield $LY$ is determined for each channel as $LY_i=\mathit{MIP}_i/\mathit{Gain}_i$. The light yield is a property of the system of scintillator tile and SiPM, while the influence of the ASIC and the SiPM gain is removed. The mean light yield of $13.8$  pixels per MIP (see figure~\ref{fig:LY}) is slightly lower than the planned value of $15$, but well within the range for a stable and efficient detector. The light yield has a spread of $1.6$ pixels per MIP, corresponding to $12\%$. The fact that the light yield spread is smaller than the spread of the MIP conversion factors shows that the spread of the ASIC gain and of the SiPM gain contribute considerably to the MIP spread. In the light yield distributions within a layer (figure~\ref{fig:LYLayer}), no clear  pattern of the ASICs is visible. 

\subsection{Time Calibration}

The measured hit times in units of TDC bins are converted to nanoseconds, according to 
\begin{equation}
    T_i{\rm [ns]} = \mathit{TDCSlope}_i{\rm \left[\frac{ns}{TDC}\right]} \times T_i{\rm [TDC]} + \mathit{TDCOffset}_i{\rm [ns]} - T_\mathit{trigger}{\rm [ns]}.
\end{equation}
%\todo{make equation readable}
In this equation $\mathit{TDCSlope}_i{\rm [\frac{ns}{TDC}]}$ is the slope of the linear TDC voltage ramp in units of ${\rm \frac{ns}{TDC}}$ for cell $i$, $\mathit{TDCOffset}_i{\rm [ns]}$ is the corresponding offset of the TDC voltage ramp in ${\rm [ns]}$, $T_i{\rm [TDC]}$ is the measured hit time in units of ${\rm TDC}$ and $T_\mathit{trigger}{\rm [ns]}$ is the trigger time measured by the BIF. The slope and offset parameters are obtained in the time calibration procedure by plotting the trigger time against the TDC values of the hits and fitting a linear function to the data. As explained in section ~\ref{sub:DAQ}, the TDC ramp switches direction from one bunch crossing to the next, so the calibration has to be done for both directions. Since most of the measurements have been performed with a $250\,{\rm kHz}$ bunch clock, the time calibration is discussed for this setting.

Similar to the MIP calibration, minimum ionizing muons covering the front face of the AHCAL are used for calibration. Within one event, all measured hit times of a muon traversing the detector are related to a single trigger time, therefore the propagation time of the muon is calibrated out. The slope of the TDC is common to all channels on the same ASIC, while the offsets are obtained for every memory cell to account for tolerances and signal propagation within the ASIC. Additionally, the offsets are shifted depending on the direction of the TDC slope. Hence, the calibration consists of two slope values per ASIC and two offset values per memory cell.

The calibration requires at least $30$ hits per memory cell to perform the linear fit. With this procedure, consistent slope and offset values throughout the AHCAL are obtained and about $93\%$ of the channels are calibrated. The resulting slope parameters lie around $\pm 1.2\,{\rm ns/TDC}$ with a spread of $10\%$ and $8\%$, depending on the bunch crossing parity. 
%No significant accumulation of outliers is observed for the calibrated channels.
After the conversion of TDC bins to nanoseconds, a time walk correction is applied to account for energy-signal dependent shifts of the hit time. 
In the analysis of the time resolution for MIP tracks, the events in a range of $500\,{\rm ns}$ around the transition to the next bunch crossing are rejected to avoid non-linearities in the TDC ramp. Furthermore, only events with one hit per layer are accepted since the time measurement gets distorted with rising ASIC occupancy, an effect observed already for previous versions of the ASIC~\cite{ref:PhDThesisBrianne}.

%% file: testbeam.tex
\section{Test Beam Experience}
\label{sub:testbeam}

\subsection{Test Beam Setup at CERN}
\label{sub:SPSTestbeam}
%\todo{move section later?}
The AHCAL technological prototype was operated in three beam test periods in the H2 beam line at the CERN Super Proton Synchrotron beam test facility during the year 2018. 

\begin{figure}[htb]
  \begin{subfigure}[T]{0.49\textwidth}
     \includegraphics[trim=100 100 1400 1100, clip,width=\linewidth]{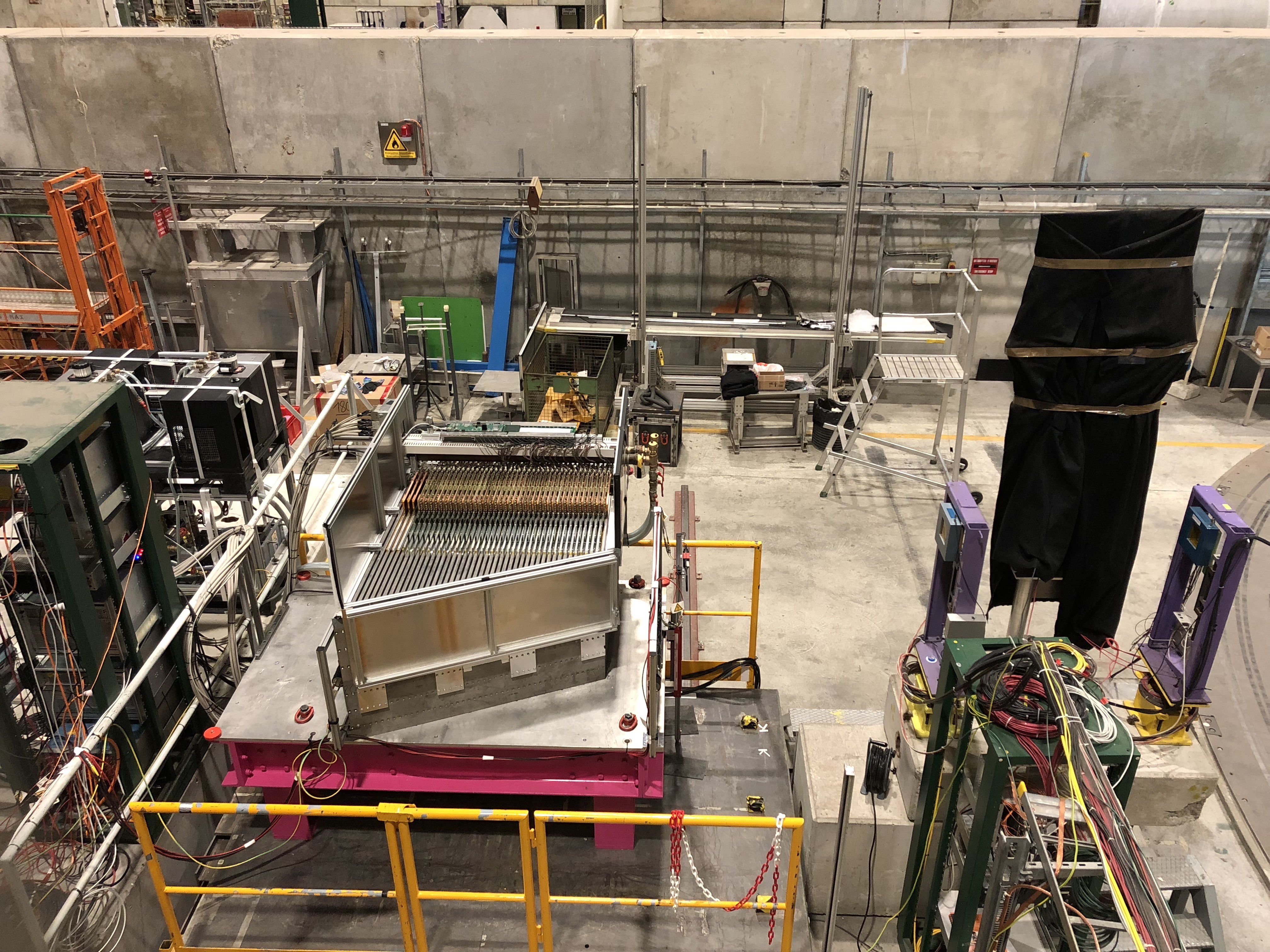}
     \caption{}\label{fig:StackAtCERN}
  \end{subfigure}%
  \hfill
  \begin{subfigure}[T]{0.49\textwidth}
     \includegraphics[width=\linewidth]{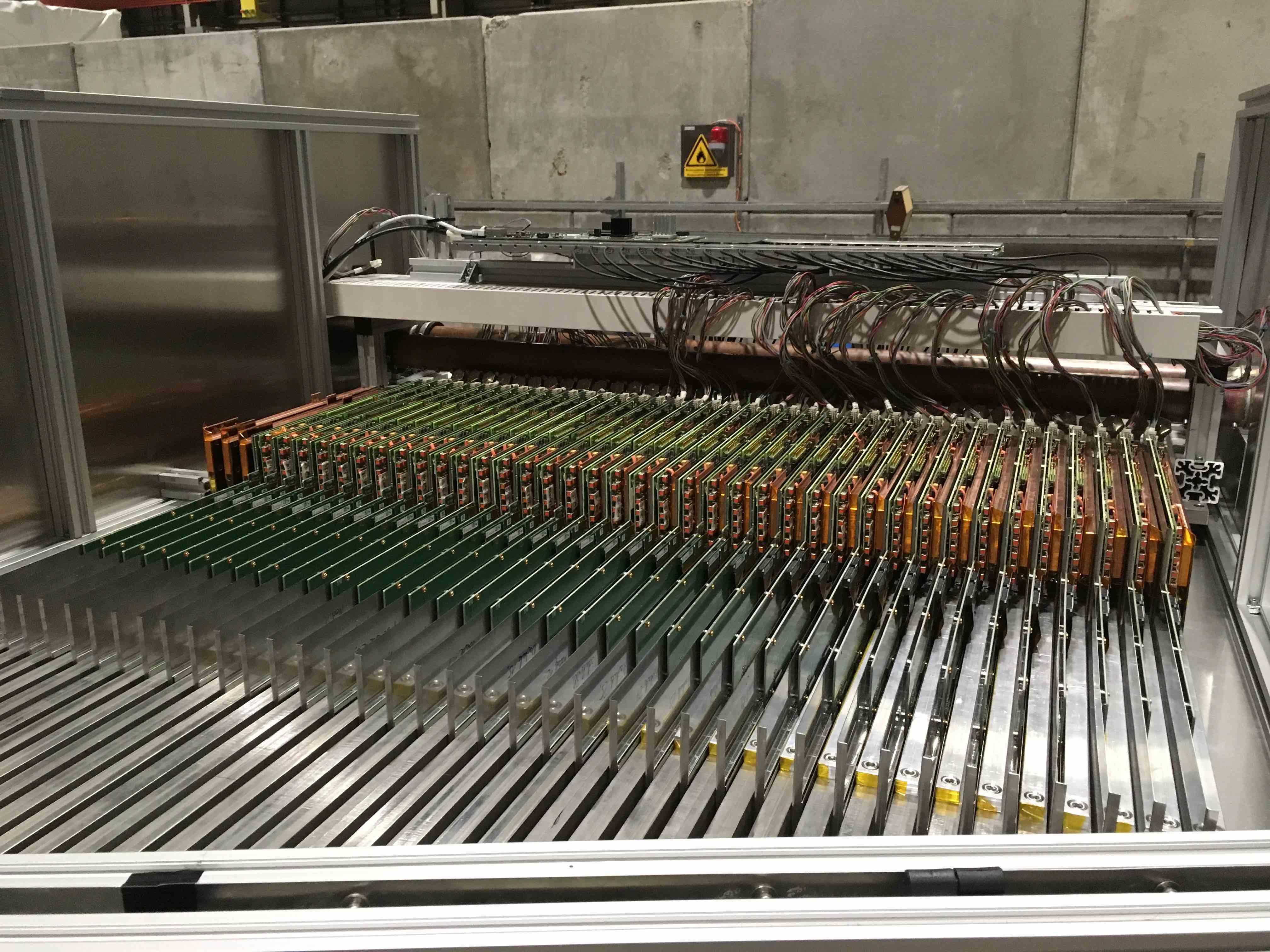}
     \caption{}\label{fig:LayersInStack}
  \end{subfigure}%
  \caption{Photos of the assembled prototype: (\subref{fig:StackAtCERN}) Overview of the setup on the x-y-table in the H2 beam line at the CERN Super Proton Synchrotron.   (\subref{fig:LayersInStack}) Close-up of the interface boards sticking out of the absorber.}
    \label{fig:AssembledPrototype}
\end{figure}

During the first two weeks of beam time in May 2018, the prototype was equipped with $38$ modules with $3 \times 3\,{\rm cm}^2$ scintillator tiles produced according to the scalable construction procedures described in section~\ref{sub:Construction}. They were installed in the first $38$ gaps of the absorber structure. The coincidence of two external trigger scintillators in the beam line was used as a reference for event building. Four delay wire chambers provided a track position. The prototype was installed on a table movable in both directions perpendicular to the beam direction (x-y table) to allow a scan of the complete detector with muons. Photos of the setup are shown in figure~\ref{fig:AssembledPrototype}. It was operated with and without power pulsing, enabling a direct comparison between the two modes. All data were taken with temperature compensation (see section~\ref{TemperatureCompensation}). The prototype was exposed to negative muon, electron and negative pion beams. In order to monitor the stability of the MIP calibration, muons were collected at the beginning and the end of the data taking, with energies of $40$ and $120\,{\rm GeV}$. Electrons with energies between $10$ and $100\,{\rm GeV}$ and pions between $10$ and $160\,{\rm GeV}$ were measured. In order to determine the gain inter-calibration factors, for some runs the hit time information was dropped, and the amplitude was read out in high gain and in low gain modes. 

For the second beam period of one week in June 2018, one module with $6 \times 6\,{\rm cm}^2$ scintillator tiles was added. It was installed in the $38^{\rm th}$ absorber gap, and the module previously installed there was moved to the $41^{\rm st}$ gap. In addition, the setup was extended by a {\it pre-shower} layer consisting of one HBU in front of the prototype, and a {\it tail catcher} behind it. The tail catcher consisted of twelve layers of one HBU each between absorbers of $7.4\,{\rm cm}$ thick steel. These HBUs were equipped with older SiPMs and some of them with a different tile geometry. The tail catcher was read out with a second wing-LDA which was fully integrated into the AHCAL DAQ. The complete setup was installed on the x-y-table. Since the experience in May had been positive, all data were taken with power pulsing as well as temperature compensation. Muons of $40\,{\rm GeV}$ energy, positrons between $10$ and $100\,{\rm GeV}$ and negative pions between $10$ and $200\,{\rm GeV}$ were measured. Due to improved beam steering, impurities observed in May for electrons and low energy pions could be reduced.

In October 2018, the AHCAL prototype with $39$ layers was operated together with the prototype for the silicon part of the CMS HGCAL~\cite{ref:HGCALprototype,ref:HGCALprototypeDAQ} for two weeks. Due to the size of the complete setup, it had to be installed on a fixed platform. Since the AHCAL prototype was placed behind the HGCAL prototype with a thickness of about $5$ nuclear interaction lengths, only muons and the tails of hadron showers reached the AHCAL. The synchronisation of the two prototypes was implemented in EUDAQ2. The data taking rate was limited by the HGCAL prototype to about $50$ events per second, resulting in about $2.5$ million events in the AHCAL prototype.

During the five weeks of beam time and additional time of LED and cosmics data taking before, the DAQ operated reliably. In total $\sim 93$ million beam events were collected.

\subsection{Detector Stability}

The stability of the detector operation was monitored in several ways: during a testbeam period, short LED runs were recorded daily, allowing the stability of the pedestal as well as the gain to be monitored on short time scales. Long LED runs in inter-calibration mode as well as muon data runs were recorded in all three testbeam periods, such that the long-term stability of the gain, the inter-calibration factor and the MIP calibration factor could also be monitored.

The pedestal values were found to be stable on short and long time scales, they are consistent between all three testbeam periods. The stability of the gain, the inter-calibration factors and the MIP calibration factors between the beam test periods was excellent: 
the gain showed a mean shift of significantly less than a percent and a channel-by-channel spread of the shift of around $1\%$; 
%A larger shift observed for one module could be traced to a broken channel of the MPOD power supply. 
the inter-calibration factors showed a mean ratio of $1.00$ and a channel-to-channel spread of $2.3\%$;
the MIP calibration factors showed a mean shift of $1\%$ with a spread of $2.7\%$ between May and June 2018.

\begin{figure}[htb]
  \begin{subfigure}[T]{0.49\textwidth}
     \includegraphics[width=\linewidth]{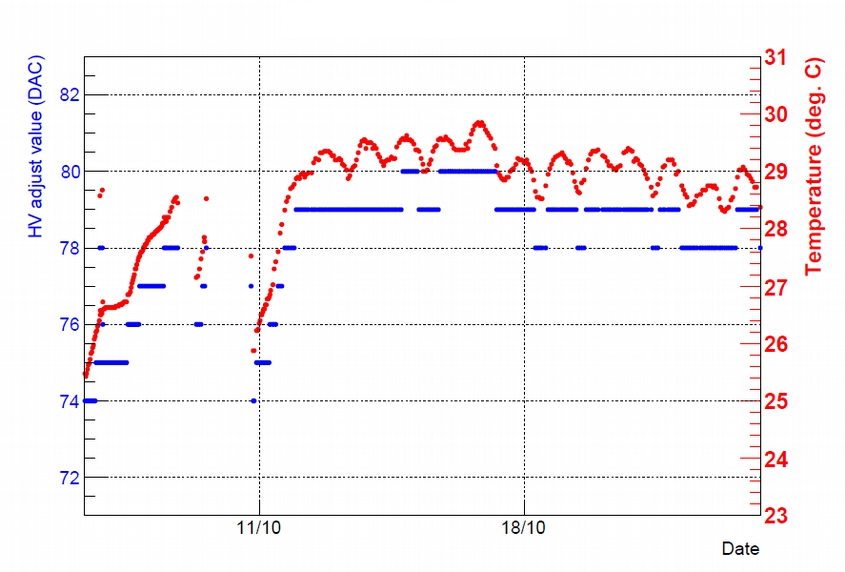}
     \caption{}\label{fig:GainComp_HV}
  \end{subfigure}%
  \hfill
  \begin{subfigure}[T]{0.49\textwidth}
     \includegraphics[width=\linewidth]{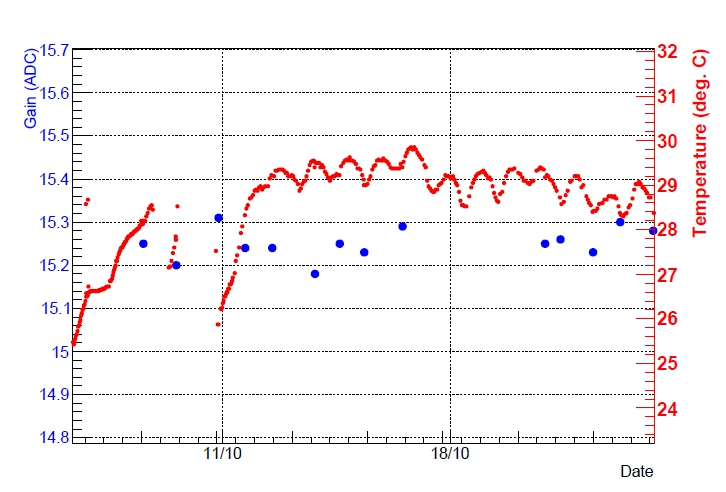}
     \caption{}\label{fig:GainComp_Gain}
  \end{subfigure}%
  \caption{(\subref{fig:GainComp_HV}) Temperature (red) and SiPM bias adjustment (blue) values and (\subref{fig:GainComp_Gain}) Temperature (red) and Gain (blue) values during the October testbeam period. }
    \label{fig:TempComp}
\end{figure}

The daily LED runs were used to monitor the dependence of the gain on the temperature. Since the electronics within the absorber structure is not cooled directly, the average temperature of the prototype varies slowly with time. After a warm-up phase in the beginning, a day-night variation of the order of one degree is observed. The results for the temperature dependence of the gain~\cite{ref:TempComp} are shown in figure~\ref{fig:TempComp} for the beam period in October 2018. All gain values agree with each other to better than $1\%$, which is the expected accuracy due to the gain determination procedure and the granularity of the SiPM bias voltage adjustment.

\subsection{Comparison of MIP Calibrations}

\begin{figure}[htb]
  \begin{subfigure}[T]{0.49\textwidth}
     \includegraphics[width=\linewidth]{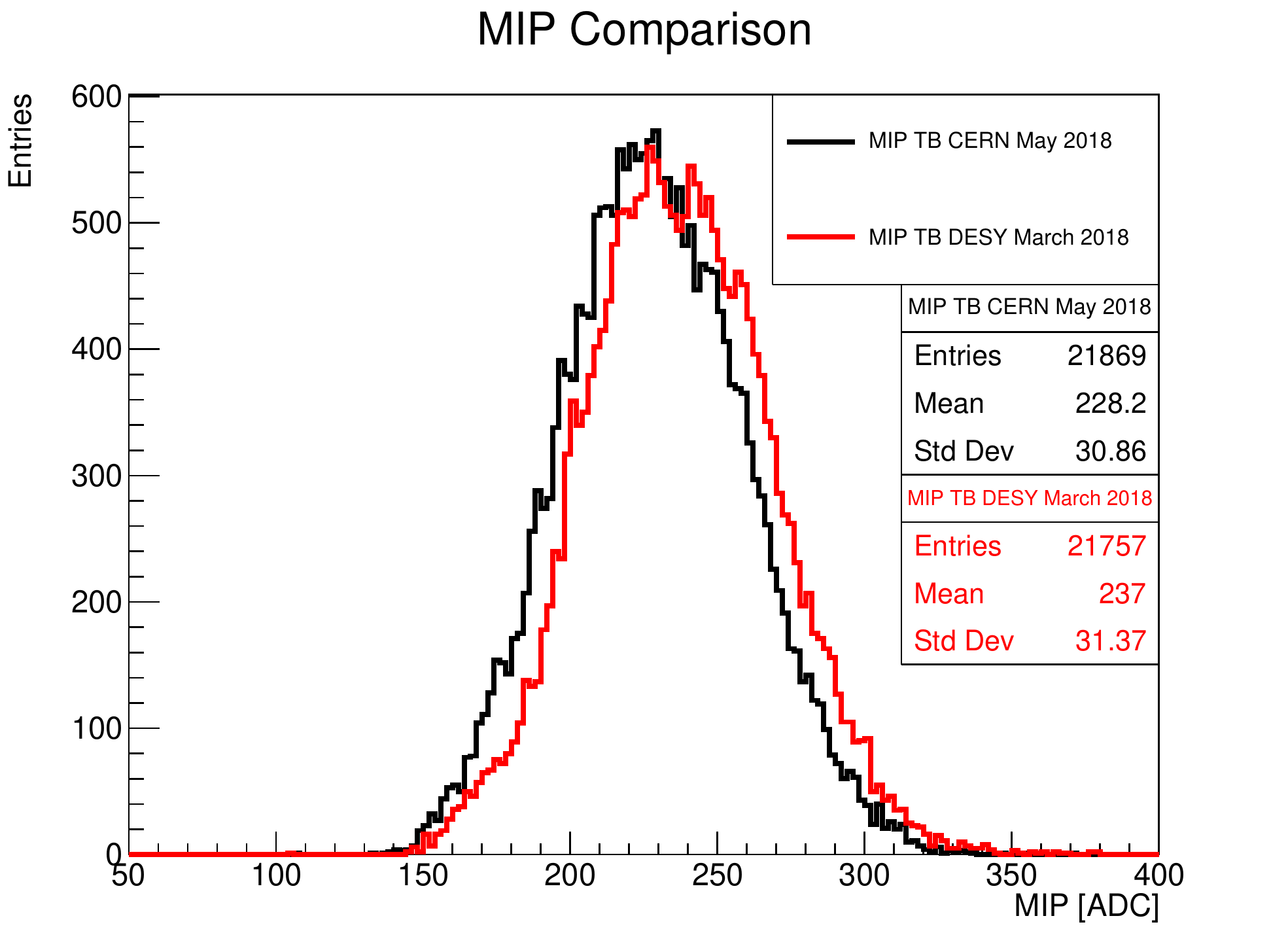}
     \caption{}\label{fig:mip_comparison_absolute}
  \end{subfigure}%
  \hfill
  \begin{subfigure}[T]{0.49\textwidth}
     \includegraphics[width=\linewidth, height=5.5cm]{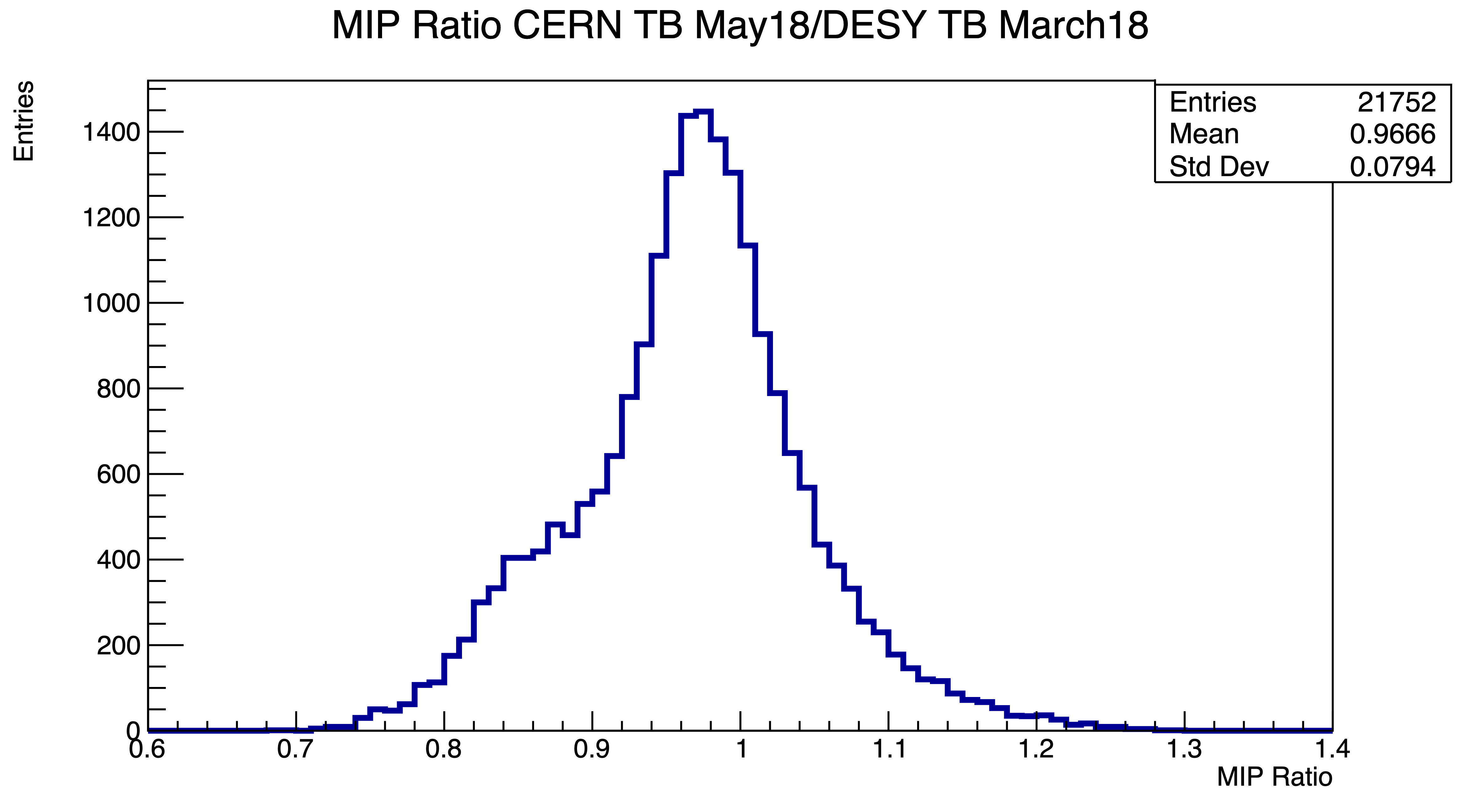}
     \caption{}\label{fig:mip_comparison_ratio}
  \end{subfigure}%
  \caption{(\subref{fig:mip_comparison_absolute}) Comparison of the MIP calibration factors determined in the DESY beam test during commissioning (red) and during the first beam test at CERN (black). (\subref{fig:mip_comparison_ratio}) Ratio of the MIP calibration factors determined in the two beam tests. }
    \label{fig:mip_comparison}
\end{figure}

During the construction and commissioning phase as well as during the testbeam periods, the MIP calibration factors were determined with several methods: after the tile assembly the individual HBUs were calibrated with cosmic rays; after the module assembly several modules (without absorber) were calibrated with cosmic rays and with non-showering electrons in the DESY testbeam facility; and during the testbeam at CERN beam muons were recorded. The MIP calibration factors from the DESY and the CERN beam tests are directly comparable, since they were recorded with consistent detector settings. The comparison is shown in figure~\ref{fig:mip_comparison}. On average, the  calibration factors determined from the electron beam at DESY are about $3\%$ higher, with a spread of about $8\%$. This demonstrates that the extraction procedure of the MIP calibration factors is robust and can also be applied for non-showering electron data, and that the values determined during the detector commissioning from electron data provide a good initial calibration for the prototype. 

\begin{figure}[htb]
%  \begin{subfigure}[T]{0.49\textwidth}
%     \includegraphics[width=\linewidth]{./figures/LY_comparison_cern_may18_desy_march18.pdf}
%     \caption{}\label{fig:LY_comparison_absolute}
%  \end{subfigure}%
%  \hfill
%  \begin{subfigure}[T]{0.49\textwidth}
%     \includegraphics[width=\linewidth]{./figures/LY_ratio_cern_may18_desy_march18.pdf}
%     \caption{}\label{fig:LY_comparison_ratio}
%  \end{subfigure}%
%  \\
  \begin{subfigure}[T]{0.49\textwidth}
     \includegraphics[width=\linewidth]{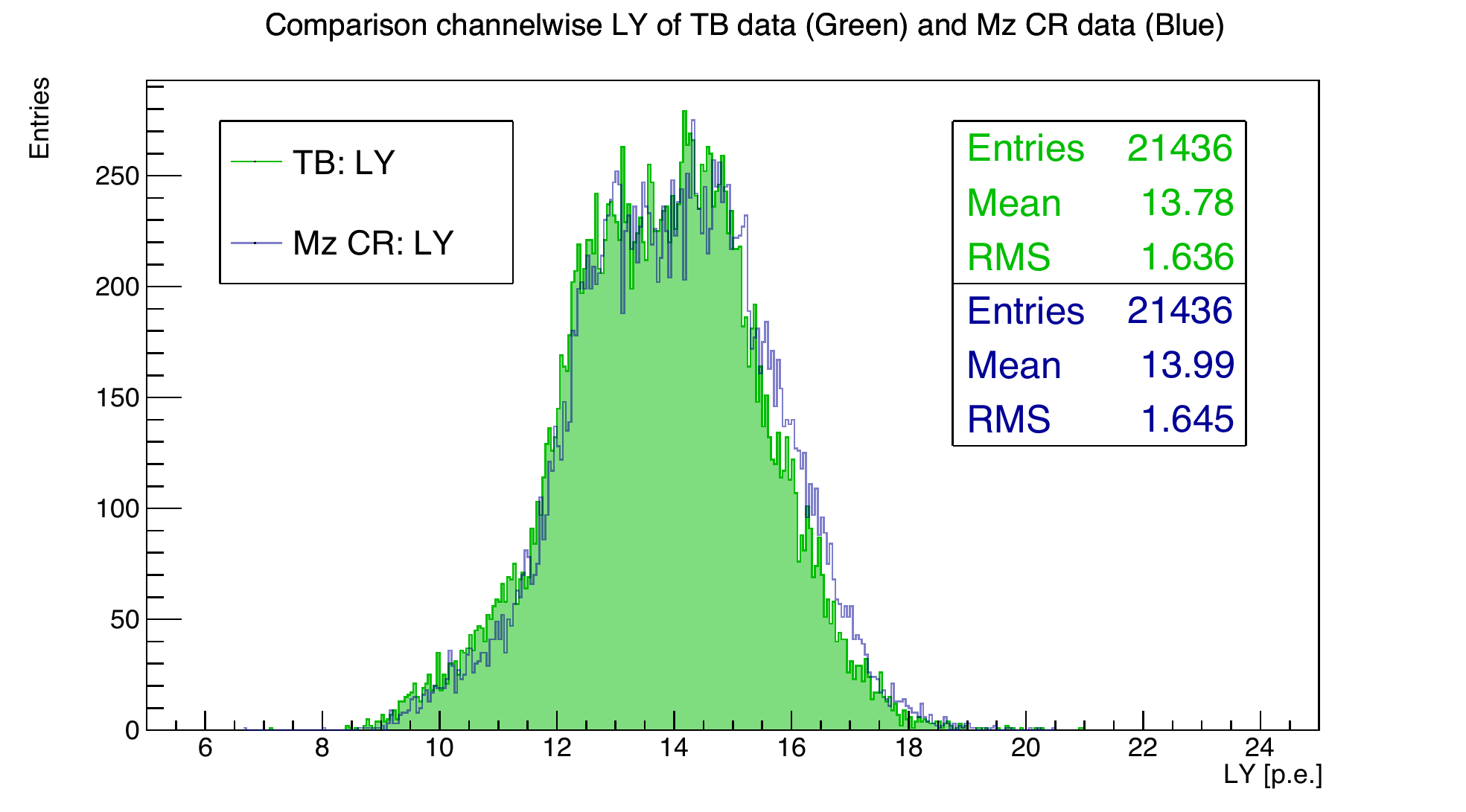}
     \caption{}\label{fig:LY_comparison_absolute}
  \end{subfigure}%
  \hfill
  \begin{subfigure}[T]{0.49\textwidth}
     \includegraphics[width=\linewidth]{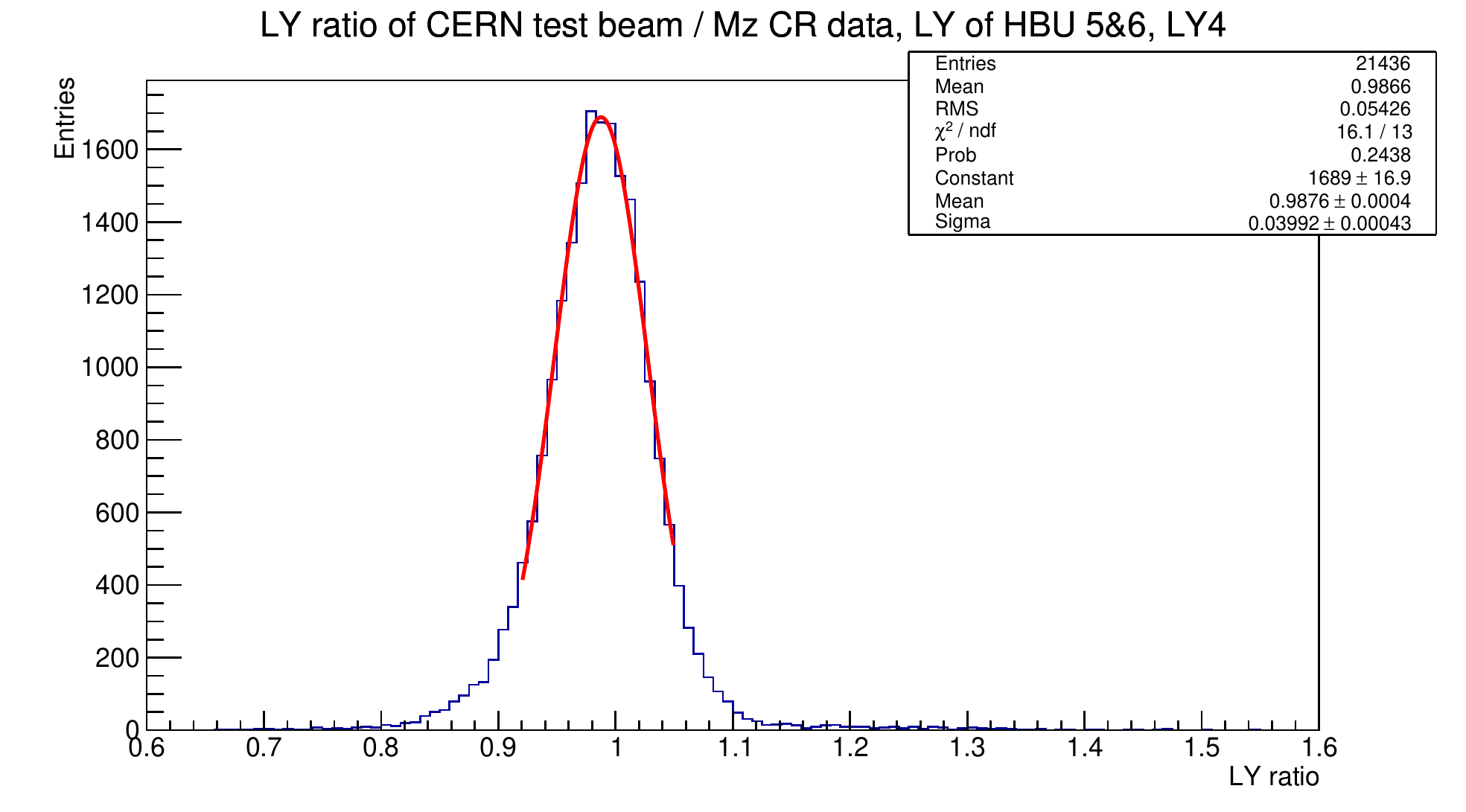}
     \caption{}\label{fig:LY_comparison_ratio}
  \end{subfigure}%
    \caption{(\subref{fig:LY_comparison_absolute}) Comparison of the light yield determined with cosmic rays after tile assembly (Mainz Cosmic Ray test stand, blue) and with muon data recorded during the first beam test at CERN (TestBeam, green). (\subref{fig:LY_comparison_ratio}) Ratio of the light yield values. The blue histogram shows the data, while the red line indicates a Gaussian fit to the central part of the distribution.}
    \label{fig:LY_comparison}
\end{figure}

During the tile assembly, the cosmic rays were recorded with a different setting of the ASIC preamplifier capacitances from the later beam data. Therefore, only the light yield values are directly comparable. The comparison presented in figure~\ref{fig:LY_comparison} shows good agreement, with a mean shift of less than $2\%$ and a spread of about $5\%$. 

\subsection{Time Resolution}
\begin{figure}[tbh]
  \begin{subfigure}[T]{0.48\textwidth}
     \includegraphics[width=\linewidth]{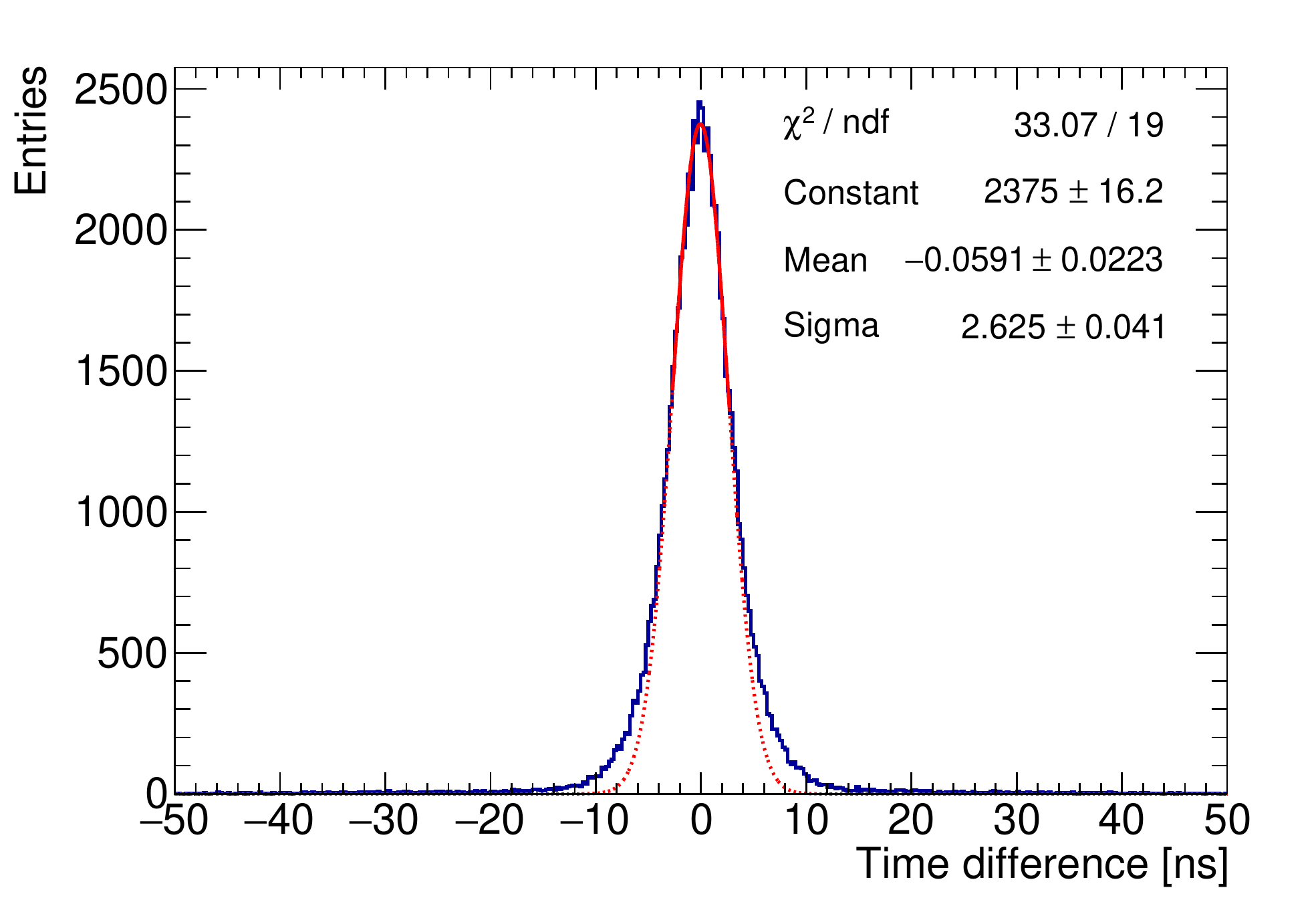}
     \caption{}\label{fig:timediffTB}
  \end{subfigure}%
  \hfill
  \begin{subfigure}[T]{0.50\textwidth}
     \includegraphics[width=\linewidth]{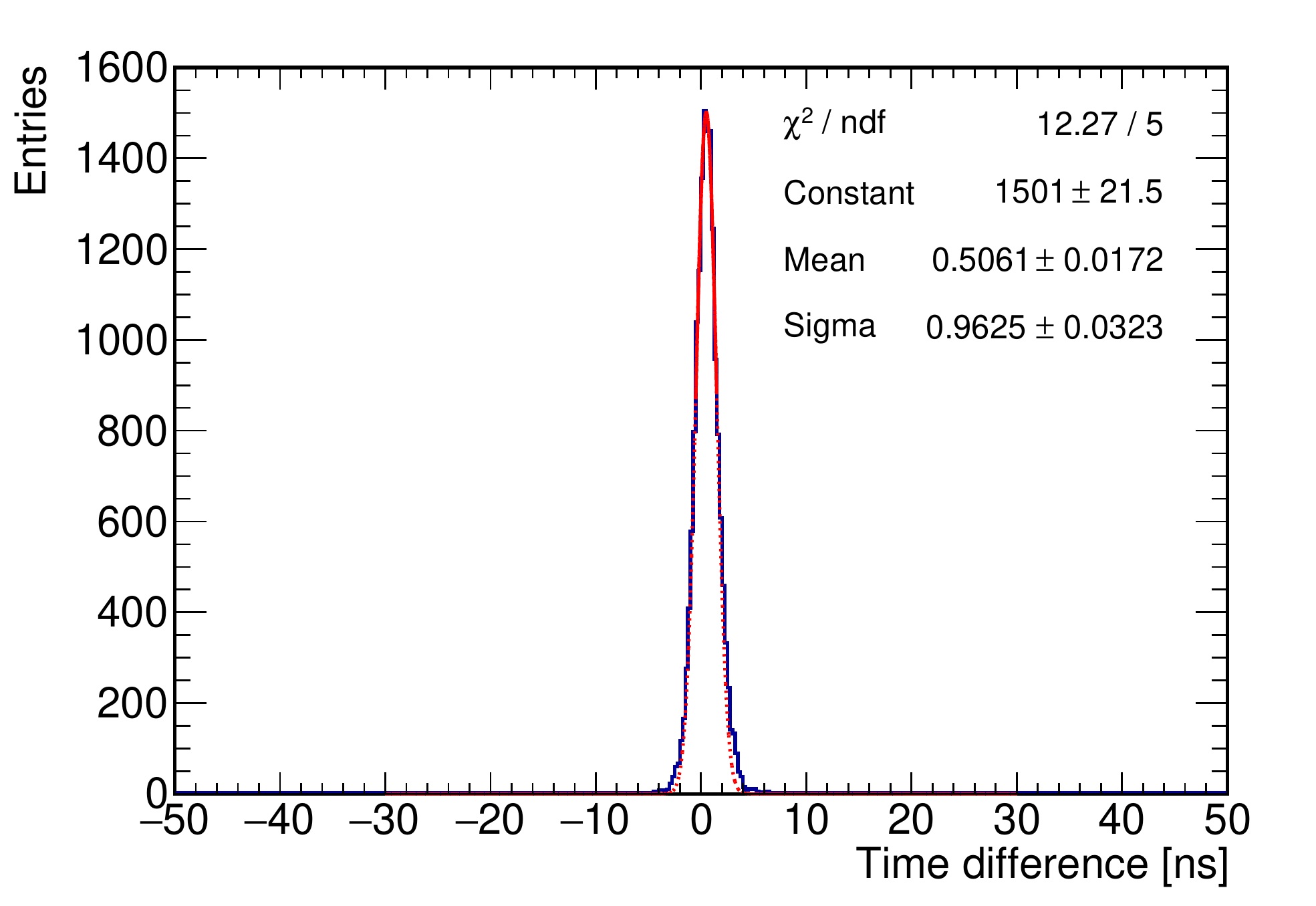}
     \caption{}\label{fig:timediffILC}
  \end{subfigure}%
  \caption{(\subref{fig:timediffTB})  Time difference of hits in subsequent layers with slow bunch clock (testbeam mode). (\subref{fig:timediffILC}) Time difference of hits in subsequent layers with fast bunch clock (ILC mode). The blue histograms show the data, while the red lines indicate  Gaussian fits (full line: fit range of mean $\pm\, 1 \sigma$, dotted line: extrapolation). The fit parameters are given in the figure.}
    \label{fig:TimeResolutions}
\end{figure}

In order to measure the time resolution of a single channel, without the influence of the time reference used in the calibration of the hit time measurement, the time difference between two hits in pairs of subsequent layers is studied. 
The resulting distribution for the testbeam bunch clock frequency of $250\,{\rm kHz}$ (see section~\ref{sub:DAQ}) is shown in figure~\ref{fig:timediffTB}. The Gaussian fitted to the data has a width of $2.6\,{\rm ns}$, corresponding to a time resolution of $1.8\,{\rm ns}$ for the individual channel. 

The hit time calibration procedure described in the previous section has also been applied for some test runs taken with the ILC mode bunch clock frequency of $5\,{\rm MHz}$ (see section~\ref{sub:DAQ}). The parameters for the TDC slope are $\pm 0.06\,{\rm ns/TDC}$, in agreement with a $20$ times faster bunch clock frequency compared to the testbeam clock frequency. The corresponding distribution of the time difference for hit pairs is shown in figure~\ref{fig:timediffILC}, resulting in a single channel time resolution with this clock setting of $0.68\,{\rm ns}$ (derived from the core width of the time difference between two channels of $0.96\,{\rm ns}$ and assuming the same resolution for both channels). The performance of the time measurement does not change significantly when running in power pulsing mode, regardless of the bunch clock frequency. Since this time resolution is determined from channels on different ASICs in different layers, it includes all detector effects: the light production and propagation in the scintillator, the signal formation in the SiPM, ASIC effects as well as effects from the distribution of the clock signal to the ASICs. It demonstrates that the design goal of a time resolution of $1\,{\rm ns}$ has been reached.

\subsection{Event Displays}

\begin{figure}[htb]
  \begin{subfigure}[T]{0.49\textwidth}
     \includegraphics[trim=70 40 700 197,clip,width=\linewidth]{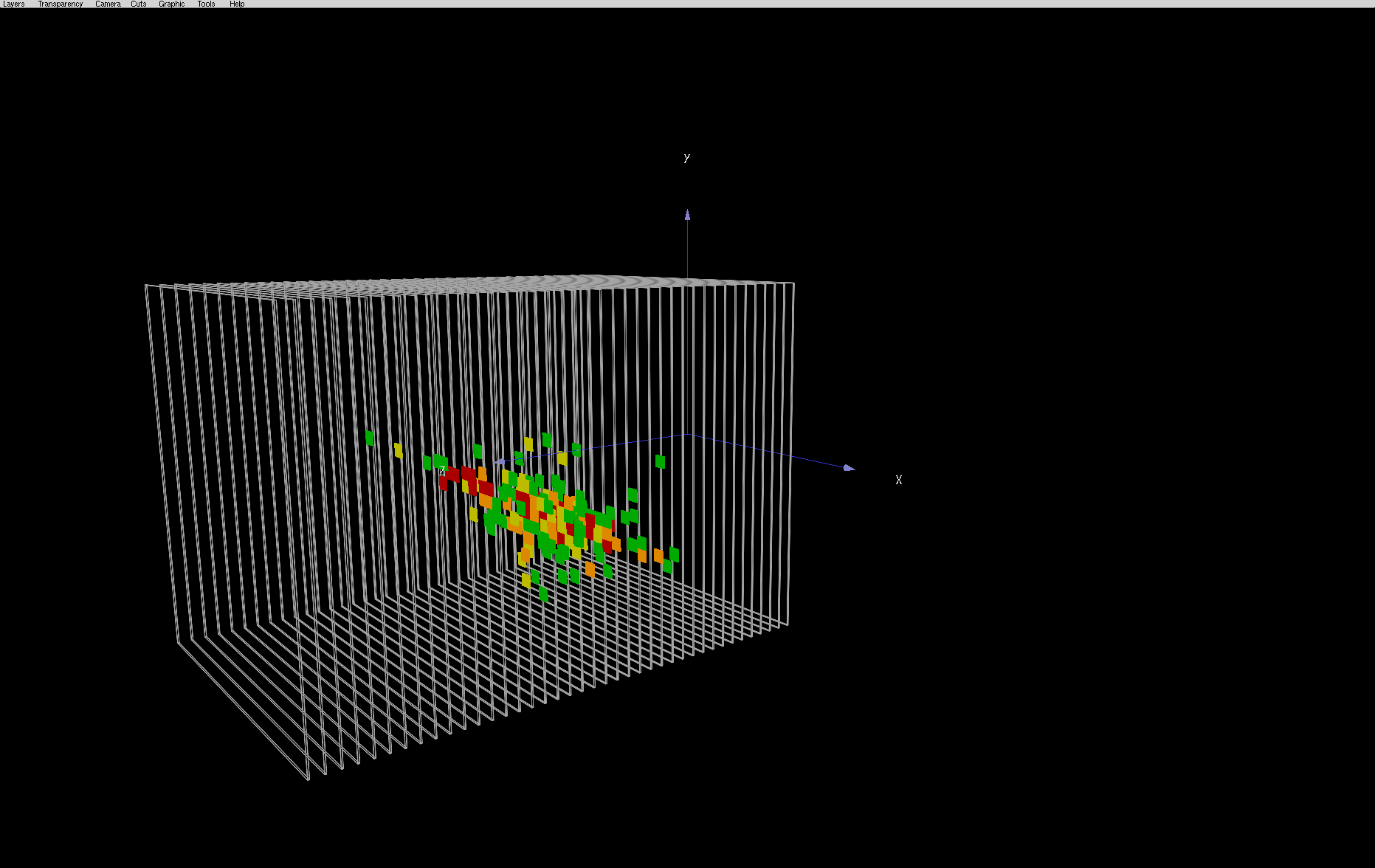}
     \caption{}\label{fig:EventCosmic}
  \end{subfigure}%
  \hfill
  \begin{subfigure}[T]{0.49\textwidth}
     \includegraphics[width=\linewidth]{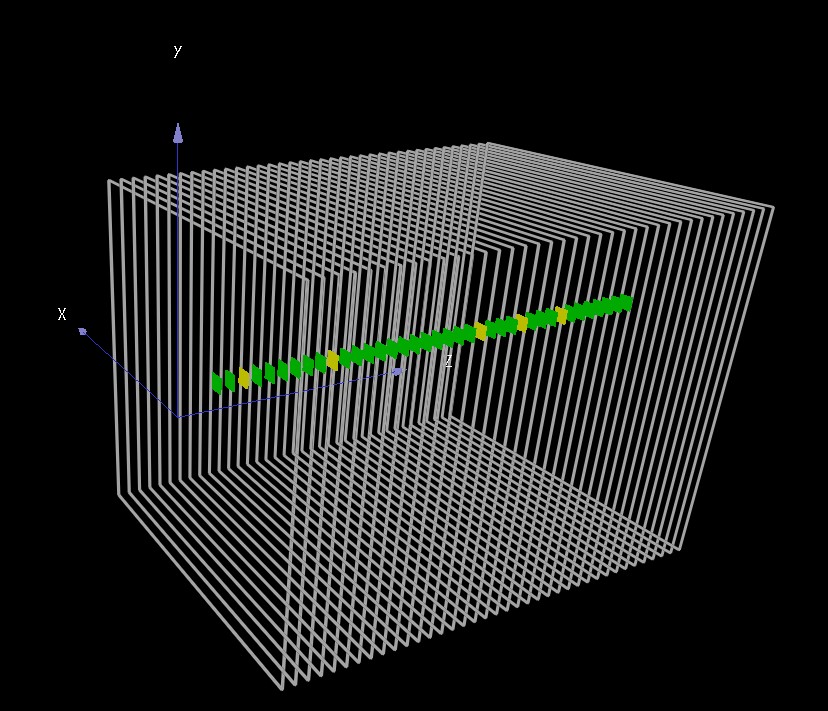}
     \caption{}\label{fig:EventMuon}
  \end{subfigure}%
  \caption{Examples of events recorded with the AHCAL technological prototype: (\subref{fig:EventCosmic}) cosmic ray muon interacting inside the detector,  (\subref{fig:EventMuon}) beam muon. The colours show hits in several amplitude ranges: green hits are in the range $0.5-1.65$ MIP, yellow in the range $1.65-2.9$ MIP, orange in the range $2.9-5.4$ MIP, and red above $5.4$ MIP.}
    \label{fig:Events1}
% if (var < 0.5) { //Not used due to 0.5 MIP cut in reco
%              color = 0x444444 ;//grey
%            }
%            else if ( (var >= 0.5) && (var < 1.65) ) {
%              color = 0x00aa00 ;//green
%              layer += 1;
%            }
%            else if ( (var >= 1.65) && (var < 2.9) ) {
%              color = 0xbbbb00;//kaki
%             layer += 2;
%            }
%            else if ( (var >= 2.9) && (var < 5.4) ) {
%              color = 0xdd8800;//gold
%              layer += 3;
%            }
%            else if ( var >= 5.4 ) {
%              color = 0xaa0000 ;//red
%              layer += 4;
%            }    
\end{figure}

\begin{figure}[htb]
  \begin{subfigure}[T]{0.49\textwidth}
     \includegraphics[trim=0 0 15 0,clip,width=\linewidth]{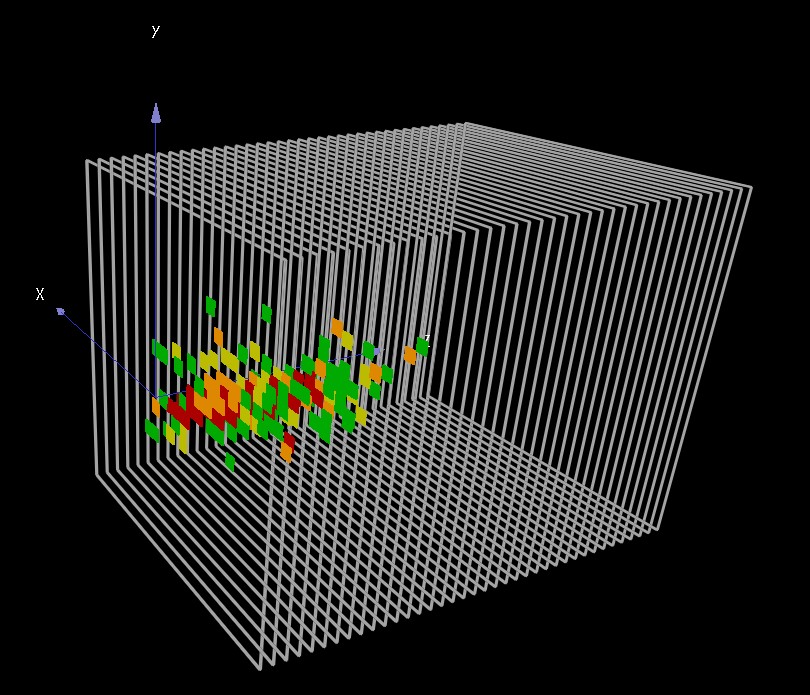}
     \caption{}\label{fig:EventElectron}
  \end{subfigure}%
  \hfill
  \begin{subfigure}[T]{0.49\textwidth}
     \includegraphics[width=\linewidth]{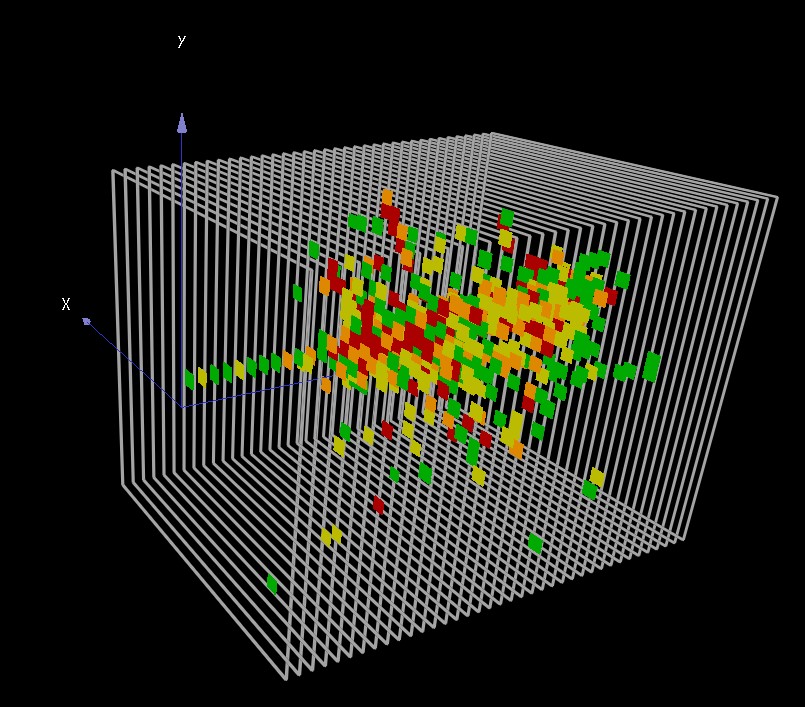}
     \caption{}\label{fig:EventPion}
  \end{subfigure}%
  \caption{Examples of events recorded with the AHCAL Technological Prototype: (\subref{fig:EventElectron}) $50\,{\rm GeV}$ electron,  (\subref{fig:EventPion}) $80\,{\rm GeV}$ pion. For the explanation of the colours see figure~\ref{fig:Events1}.}
    \label{fig:Events2}
\end{figure}

%In order to demonstrate that capabilities the AHCAL technological prototype, 
Figures~\ref{fig:Events1} and~\ref{fig:Events2} show three-dimensional views of events: figure~\ref{fig:Events1} shows muons, one in a cosmic ray event recorded during the commissioning, causing a shower, and a typical beam muon. Figure~\ref{fig:Events2} shows an electron shower and a pion showering in layer $\approx 10$ after leaving a track-like signature in the layers before. These displays illustrate the excellent imaging capabilities of the AHCAL.

\begin{figure}[htb]
  \begin{subfigure}[T]{0.49\textwidth}
     \includegraphics[width=\linewidth]{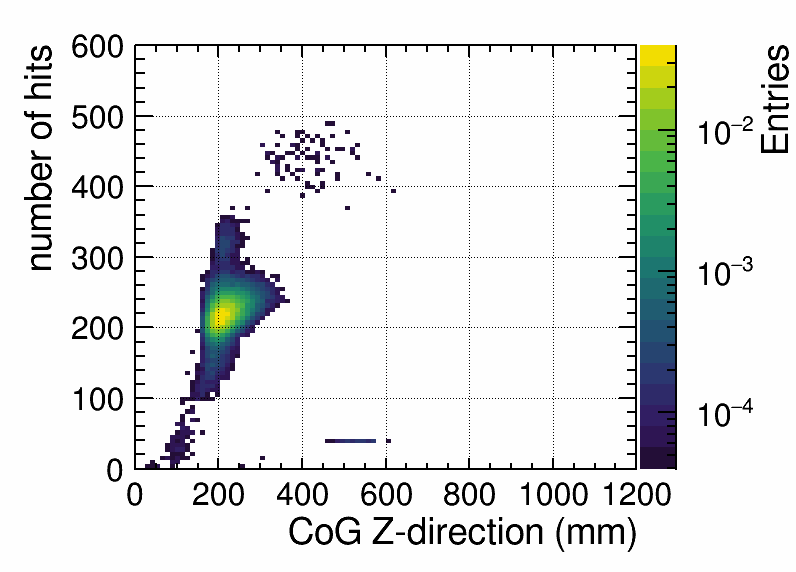}
     \caption{}\label{fig:NHitvsCOGzElectron}
  \end{subfigure}%
  \hfill
  \begin{subfigure}[T]{0.49\textwidth}
     \includegraphics[width=\linewidth]{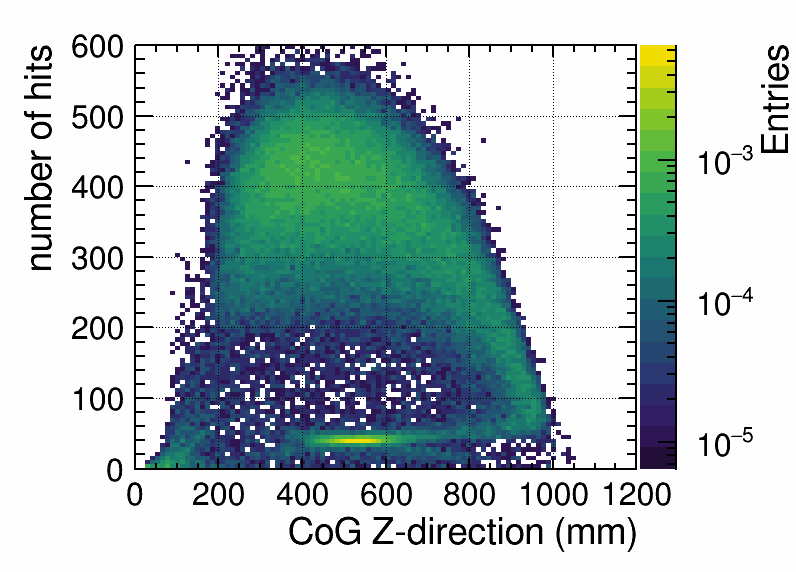}
     \caption{}\label{fig:NHitvsCOGzPion}
  \end{subfigure}%
 \caption{Distribution of the number of hits, $N_{hit}$, and the centre-of-gravity along the beam axis, $CoG_{Z}$ for (\subref{fig:NHitvsCOGzElectron}) a $60\,{\rm GeV}$ electron run and (\subref{fig:NHitvsCOGzPion}) a $60\,{\rm GeV}$ pion run with small contamination of other particle species. 
 %: electron showers show a narrow distribution in $N_{hit}$ and $CoG_{Z}$ close to (200, 200 mm), while pion showers show a wider distribution in both quantities, with (on average) more hits and deeper center-of-gravity than electrons. For pions which start to shower late in the detector, a large $CoG_{Z}$ and a reduced number of hits due to longitudinal leakage is observed.  Muons do not shower, so have a number of hits close to the number of active layers, and a center-of-gravity close to the middle of the detector, so accumulate around (39, 500 mm). 
 }
    \label{fig:NHitvsCOGz}
\end{figure}

The detailed information provided by the AHCAL can be used to monitor not only the detector stability, but also beam parameters like the purity, already during data taking. An example of a distribution used regularly for this purpose is shown in figure~\ref{fig:NHitvsCOGz} for an electron and a pion run with low contamination: for a given beam energy, electron showers show a narrow distribution in the number of hits and the center-of-gravity along the beam axis at a small detector depth, while hadron showers show a wider distribution in both quantities, with (on average) more hits and deeper centre-of-gravity than electrons. Muons usually do not shower, so have a number of hits close to the number of active layers, and a center-of-gravity close to the middle of the detector.

%\todo{get consistent pictures}
%\todo{get a nice N\_hit vs z\_cog plot}

%% file: conclusion.tex
\section{Conclusion}
We have constructed a highly granular $38$-layer analog hadron calorimeter prototype consisting of steel absorber and scintillator tiles individually read out with directly coupled SiPMs. The focus of the design of the technological prototype lies on the scalability of the detector layout to a collider detector and the scalability of the production methods. The beam tests show a reliable operation of the prototype and the feasibility of the calibration procedures for a large amount of readout channels. The detector shows very low noise and excellent uniformity of the response. After the demonstration of the physics potential of a highly granular steel-scintillator hadron calorimeter with the earlier AHCAL physics prototype, this is an important step towards the realization of the concept as hadron calorimeter in a collider detector. The individual hit time measurement with a resolution of better than $1\,{\rm ns}$ opens the way for future detailed studies of hadron shower development not only in space, but also in time. 